\documentclass[%
 reprint,
 amsmath,amssymb,
 aps,
prb,
 superscriptaddress
]{revtex4-1}

\usepackage{graphicx}
\usepackage{dcolumn}
\usepackage{bm}
\usepackage{braket}

\begin{document}

\preprint{APS/123-QED}

\title{Topological phases in a Weyl semimetal multilayer}

\author{Kazuki Yokomizo}
\affiliation{Department of Physics, Tokyo Institute of Technology, 2-12-1 Ookayama, Meguro-ku, Tokyo, 152-8551, Japan}
\author{Shuichi Murakami}%
\affiliation{Department of Physics, Tokyo Institute of Technology, 2-12-1 Ookayama, Meguro-ku, Tokyo, 152-8551, Japan}
\affiliation{TIES, Tokyo Institute of Technology, 2-12-1 Ookayama, Meguro-ku, Tokyo, 152-8551, Japan}%





\begin{abstract}
We investigate multilayers of a normal insulator and a Weyl semimetal using two models: an effective model and a lattice model. As a result, we find that the behavior of the multilayers is qualitatively different depending on the stacking direction relative to the displacement vector between the Weyl nodes. When the stacking direction is perpendicular to the displacement vector between the Weyl nodes, the system shows either the normal insulator or the Weyl semimetal phases depending on the thicknesses of the two layers. In contrast, when the stacking direction is parallel to that, the phase diagram is rich, containing the normal insulator phase, the Weyl semimetal phase and the quantum anomalous Hall phases with various values of the Chern number. As a superlattice period increases, the Chern number in the quantum anomalous Hall phases increases. Thus, one can design a quantum anomalous Hall system with various Chern numbers in a multilayer of 
a Weyl semimetal and a normal insulator. Applications to Weyl semimetal materials are discussed.

\end{abstract}

\pacs{Valid PACS appear here}
\maketitle


\section{\label{sec1} Introduction}

Recently, various topological systems are actively studied in the field of condensed matter physics. After theoretical proposals of the quantum spin Hall system \cite{PhysRevLett.95.146802,Bernevig1757,PhysRevLett.98.106803}, various topological systems have been theoretically predicted or experimentally observed in real materials. In particular, a Weyl semimetal (WSM) which belongs to a class of topological semimetals has been proposed \cite{1367-2630-9-9-356,PhysRevB.83.205101}. A WSM has Dirac cones without spin degeneracy at or near the Fermi energy. It cannot be realized unless the time-reversal symmetry or the inversion symmetry is broken. The vertex of the Dirac-cone dispersion is called a Weyl node, and it can be regarded as a monopole or an anti-monopole for the Berry curvature field in the wavevector space \cite{1367-2630-9-9-356, Berry45,volovik2003universe}. Various WSMs are predicted after these works, and as candidates of WSMs, pyrochlore iridates \cite{PhysRevB.83.205101,PhysRevB.84.075129,Y2Ir2O7}, Hg${\rm Cr}_{2}{\rm Se}_{4}$ \cite{PhysRevLett.107.186806,PhysRevLett.108.266802}, TaAs \cite{PhysRevX.5.011029}, Co$_2$Ti$X$ ($X$=Si,Ge,Sn) \cite{Co2TiX}, ZrCo$_2$Sn \cite{magHeusler}, Mo$_x$W$_{1-x}$Te$_2$ \cite{MoWTe}, Ge$_2$Sb$_2$Te$_5$ \cite{Ge2Sb2Te5},  TlBiSe$_2$ \cite{TlBiSe2}, Hg$_{1-x-y}$Cd$_x$Mn$_y$Te \cite{HgCdMnTe} in a magnetic field, and Bi$_{0.97}$Sb$_{0.03}$ in a magnetic field \cite{Bi097Sb003}have been proposed. In recent years, some WSMs has been experimentally observed, such as TaAs \cite{PhysRevX.5.031013,lv2015observation,PhysRevLett.116.096801,xu2016observation,Xu613} and NbAs \cite{xu2015discovery}.

Among theoretical works of the WSM, a multilayer of a topological insulator (TI) and a normal insulator (NI) with magnetization has been proposed to realize the WSM phase \cite{PhysRevLett.107.127205}. Moreover, a TI-NI multilayer with broken inversion symmetry by an electric field has also been proposed to show the WSM phase by tuning some parameters, for example, the thickness of the layers in the multilayer of HgTe/CdTe \cite{PhysRevB.85.035103}. In addition, physical properties of multilayers of a TI and a NI have been investigated theoretically \cite{PhysRevB.85.165110,PhysRevB.86.115133,PhysRevB.88.125105}.

In this paper, we study multilayers of a WSM and a NI, which has not been studied thus far, to the authors' knowledge. We find that the resulting phase diagrams show rich physics, including the quantum anomalous Hall (QAH) phases with various values of the Chern number. The phase diagrams are qualitatively different depending on the stacking direction of the multilayers relative to the displacement vector between the Weyl nodes. Therefore, we focus on two patterns of stacking directions, patterns A and B shown in Figs.~\ref{fig0} (b) and (c), and calculate the phase diagrams of patterns A and B separately. In the calculation, we use an effective model and a lattice model, and their results well agree. In the multilayer with pattern A, the phase diagram havs only two phases, a NI and a WSM, depending on the ratio of the thicknesses of the two layers. On the other hand, in the multilayer with pattern B, the phase diagram contains a NI, a WSM, and QAH phases, depending on the thicknesses of the two constituent layers. The Chern number of the QAH phases becomes larger and larger when the superlattice periodicity becomes larger. This result offers us a way to design QAH systems in a WSM-NI multilayer in a controlled way.

This paper is organized as follows. In Sec.~\ref{sec2}, we calculate phase diagrams of multilayers for patterns A and B from the effective model for a WSM. We also calculate them using the lattice model and compare the results of the effective model with those of the lattice model in Sec.~\ref{sec3}. Finally, we discuss the results and applications to real materials in Sec.~\ref{sec4}.

\section{\label{sec2} Multilayer from the effective model}

\subsection{\label{subsec1} Effective model for a WSM}

In our calculation on multilayers, we use the effective model for a WSM proposed in Ref.~\onlinecite{PhysRevB.89.235315}. The model describes both a WSM and a NI by changing a single parameter $m$. The Hamiltonian of the model is a $2\times2$ matrix $H({\bm k},m)$, which acts on a space consisting of a single conduction band and a single valence band. The $2\times2$ Hamiltonian $H({\bm k},m)$ is given by
\begin{equation}
H({\bm k},m)=\gamma\left(k_{x}^2-m\right)\sigma_{x}+v\left(k_{y}\sigma_{y}+k_{z}\sigma_{z}\right),
\label{eq1}
\end{equation}

\noindent where $\sigma_{i}\hspace{3pt}\left(i=1,2,3\right)$ are the Pauli matrices, and $v$ and $\gamma$ are nonzero constants. We choose them to be positive for 
simplicity. Energy eigenvalues of the system are given by
\begin{equation}
E=\pm\sqrt{\gamma^2\left(k_{x}^2-m\right)^2+v^2\left(k_{y}^2+k_{z}^2\right)}.
\label{eq2}
\end{equation}

\noindent We assume that the Fermi energy is at $E=0$. When $m<0$, the system is an insulator with the bulk gap $2\gamma\left|m\right|$. On the other hand, when $m>0$, the system is a WSM. The bulk gap closes at two Weyl nodes ${\bm k}=(\pm\sqrt{m},0,0)$, which are a monopole and an antimonopole for the Berry curvature field.

Surface Fermi arcs of a WSM are also described within this model \cite{PhysRevB.89.235315}. In order to show the surface states, we consider two types of surfaces of a slab: a top surface and a bottom surface, perpendicular to the $z$ direction. The top surface represents the $z=0$ surface where the $z<0$ region is a WSM with $m>0$ and the $z>0$ region is the vacuum with $m<0$. Similarly, the bottom surface represents the $z=0$ surface where the $z>0$ region is a WSM and the $z<0$ region is the vacuum. As shown in Ref.~\onlinecite{PhysRevB.89.235315}, surface states exist both on the top and the bottom surfaces within $-\sqrt{m}<k_{x}<\sqrt{m}$. These surface states form Fermi arcs. In particular, at $E=0$, the surface states at each surface are located at $-\sqrt{m}<k_{x}<\sqrt{m}$, $k_{y}=k_{z}=0$, and they connect the Weyl nodes projected on the $xy$ plane. 

As mentioned earlier, a WSM is realized only when either the time-reversal symmetry or the inversion symmetry is broken. Equation.~(\ref{eq1}) itself preserves the inversion symmetry but breaks the time-reversal symmetry. Meanwhile, we can also describe WSMs without inversion symmetry by combining two models described by Eq.~(\ref{eq1}), which is beyond the scope of the paper.

\begin{figure}[b]
\centering
\includegraphics[width=8.6cm,height=5.0cm]{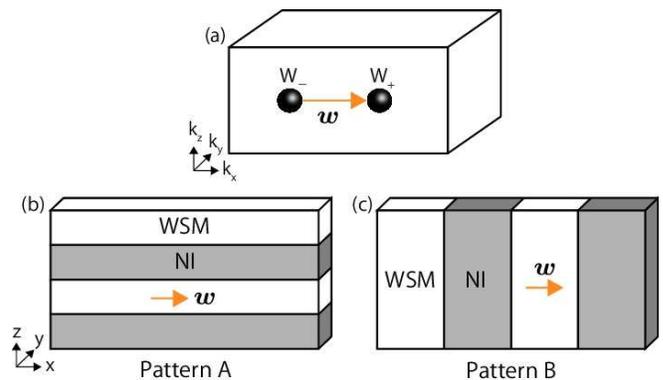}
\caption{\label{fig0} (a) Schematic diagram of the Weyl nodes, displaced along the $k_{x}$ direction in the wave-vector space. The vector ${\bm w}$ is the displacement vector between the two Weyl nodes in the ${\bm k}$ space. (b), (c) Stacking patterns of the multilayers: (b) pattern A and (c) pattern B.}
\end{figure}

The two Weyl nodes in this model are displaced along the $k_{x}$ direction in the wave-vector space [Fig.~\ref{fig0} (a)]. Therefore, in multilayers of a WSM and a NI, the stacking direction relative to the displacement vector ${\bm w}$ between the two Weyl nodes (i.e., $k_{x}$ direction), will affect their properties. Here, we study two cases for the multilayers: patterns A and B. In pattern A [Fig.~\ref{fig0} (b)], the stacking direction is perpendicular to the vector ${\bm w}$, while in pattern B [Fig.~\ref{fig0} (c)], it is parallel to the vector ${\bm w}$.

\subsection{\label{subsec2} Multilayer: Pattern A}

\begin{figure}[]
\begin{tabular}{cc}
\begin{minipage}[]{0.5\hsize}
\begin{center}
\includegraphics[width=4.1cm,height=4.0cm]{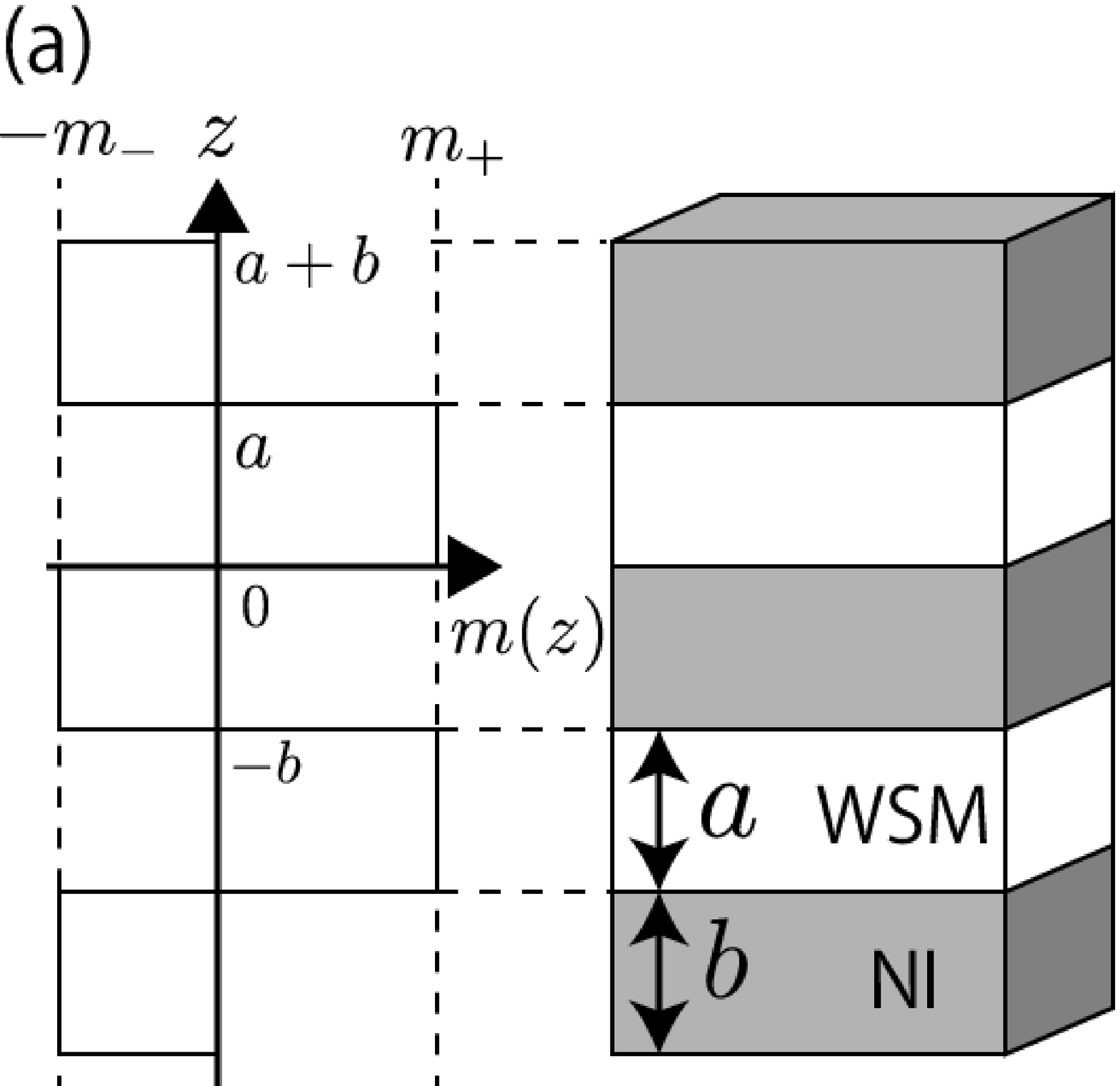}
\end{center}
\end{minipage}
\begin{minipage}[]{0.5\hsize}
\begin{center}
\includegraphics[width=3.6cm,height=4.0cm]{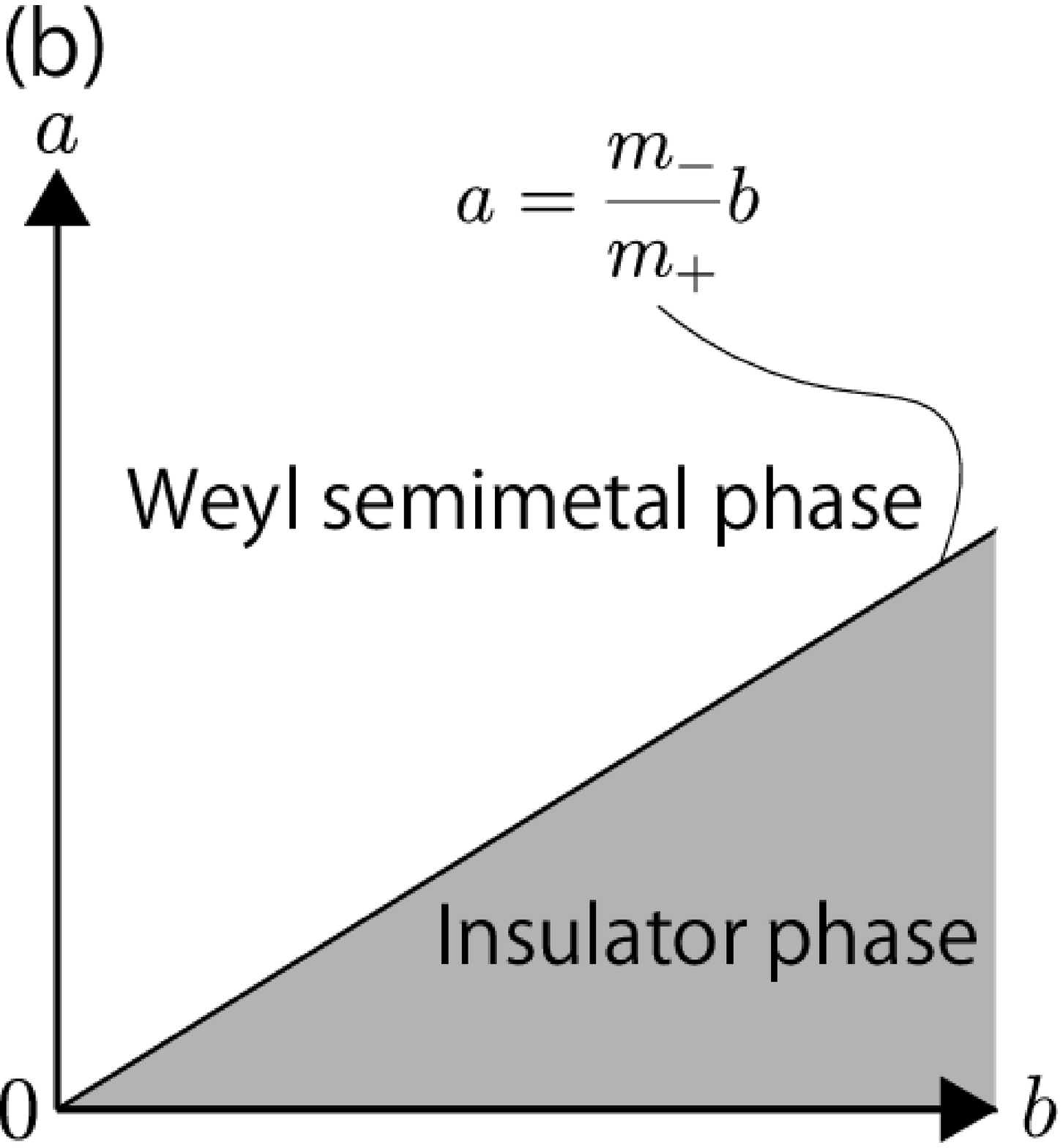}
\end{center}
\end{minipage}
\end{tabular}
\caption{\label{fig1} (a) Schematic diagram of the multilayer of pattern A and the $z$ dependence of the parameter $m(z)$. (b) Phase diagram of the WSM-NI multilayer of pattern A.}
\end{figure}

In this subsection, we discuss the WSM-NI multilayer of pattern A using the effective model (\ref{eq1}). The stacking direction is in the $z$ direction, perpendicular to the displacement vector ${\bm w}$ between the two Weyl nodes. Since the model (\ref{eq1}) represents a WSM when $m>0$ and a NI when $m<0$, the multilayer is realized by putting the control parameter $m$ to be periodic as shown in Fig.~\ref{fig1} (a), and the Hamiltonian is given by
\begin{eqnarray}
&&H=\gamma\left[k_{x}^2-m(z)\right]\sigma_{x}+v\left(k_{y}\sigma_{y}-i\partial_{z}\sigma_{z}\right), \\
&&m\left(z\right)=\left\{ \begin{array}{ll}
m_{+}  & \left(0\leq z\leq a\right), \vspace{5pt}\\
-m_{-} & \left(-b\leq z\leq0\right),
\end{array}\right. \\
&&m\left(z+\left(a+b\right)\right)=m\left(z\right),
\label{eq345}
\end{eqnarray}

\noindent where $m_{\pm}$ are positive constants. We assume the system to be infinite along the $xy$ plane. Along the $z$ axis the system has periodicity $a+b$. This model is similar to the Kr\"{o}nig-Penney model, and it can be solved similarly, by constructing plane-wave solutions in each layer and imposing proper boundary conditions at the interfaces. In the solution, a Bloch wavenumber $k_{z}$ along the $z$ direction is introduced. Detailed calculations are summarized in Appendix \ref{subsec7}.

We first investigate phases in this multilayer. When $a/b>m_{-}/m_{+}$, the gap at $E=E_{F}=0$ closes at

\begin{equation}
{\bm k}=\left(\pm\sqrt{\frac{m_{+}a-m_{-}b}{a+b}},0,0\right),
\label{eq6}
\end{equation}

\noindent and they are shown to be the Weyl nodes. Therefore, the multilayer is in the WSM phase. When $a/b<m_{-}/m_{+}$, the system is gapped. It is the NI phase because it is connected to the NI phase when $a\rightarrow0$. By increasing $a$ across the phase boundary $a/b=m_{-}/m_{+}$, the Weyl nodes are pairwise created from the origin ${\bm k}={\bm 0}$ in the wave-vector space. By varying the parameters $a$, $b$, $m_{+}$, or $m_{-}$, the Weyl nodes move along the $k_{x}$ axis from the origin ${\bm k}={\bm 0}$ in the wave-vector space within the WSM phase. Finally, in the limit of $a/b\rightarrow\infty$, the Weyl nodes converge to $(\pm\sqrt{m_{+}},0,0)$. Thus, the multilayer is in the WSM phase when the NI layer is sufficiently thin compared to the WSM layer. From these considerations, the phase diagram is as shown in Fig.~\ref{fig1} (b).

Next, we calculate the surface state of this multilayer. We consider the multilayer with semi-infinite geometry with the $z=0$ surface; the $z<0$ region is the multilayer and the $z>0$ region is the vacuum. The wave function of the surface state and its energy eigenvalue are calculated as
\begin{eqnarray}
\psi&=&\left( \begin{array}{c}
1  \\
-i
\end{array}\right)\hspace{1.5pt}{\rm e}^{-\left(\gamma/v\right)\int^z{\rm d} z\left(k_{x}^2-m\left(z\right)\right)},\hspace{3pt}E=-vk_{y}.
\label{eq7}
\end{eqnarray}

\noindent This describes the surface state only when the integral in Eq.~(\ref{eq7}) decays into the bulk; this condition is given by
\begin{equation}
\left(a+b\right)k_{x}^2-\left(m_{+}a-m_{-}b\right)<0.
\label{eq8}
\end{equation}

\noindent Hence, the surface state exists when
\begin{equation}
-\sqrt{\frac{m_{+}a-m_{-}b}{a+b}}<k_{x}<\sqrt{\frac{m_{+}a-m_{-}b}{a+b}};
\label{eq9}
\end{equation}

\noindent namely, it appears between the two Weyl nodes and forms a Fermi arc. In addition, the surface state has a velocity $(0,-v)$ along the $y$ direction from Eq.~(\ref{eq7}). From the bulk-edge correspondence, we conclude that the Chern number of the two Weyl nodes $\left(\pm\sqrt{\frac{m_{+}a-m_{-}b}{a+b}},0,0\right)$ is $\pm1$, respectively.

\subsection{\label{subsec3} Multilayer: Pattern B}

\begin{figure}[b]
\begin{center}
\includegraphics[width=5.5cm,height=4.0cm]{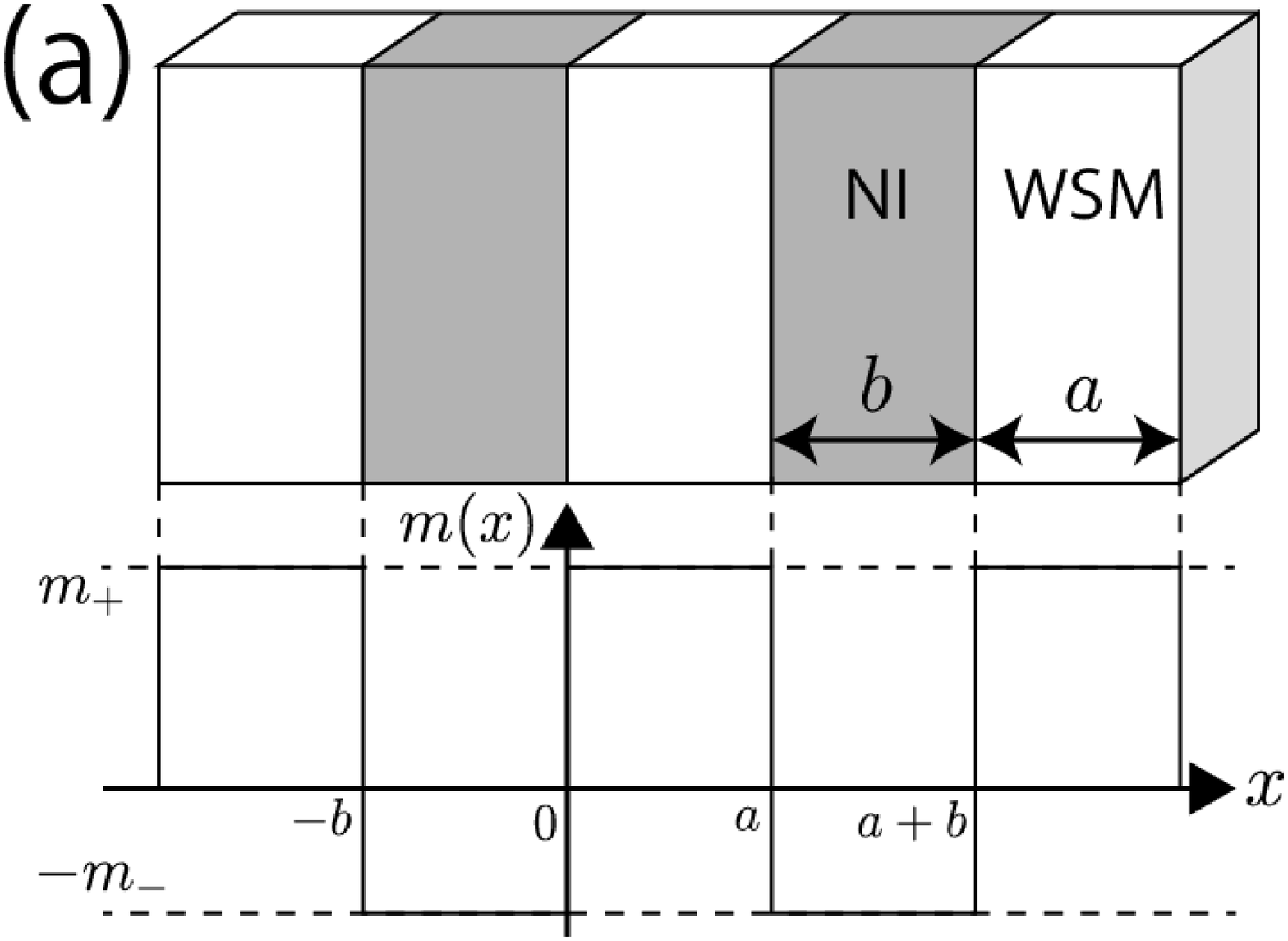}
\end{center}
\begin{tabular}{cc}
\begin{minipage}[]{0.5\hsize}
\begin{center}
\includegraphics[width=4.3cm,height=4.0cm]{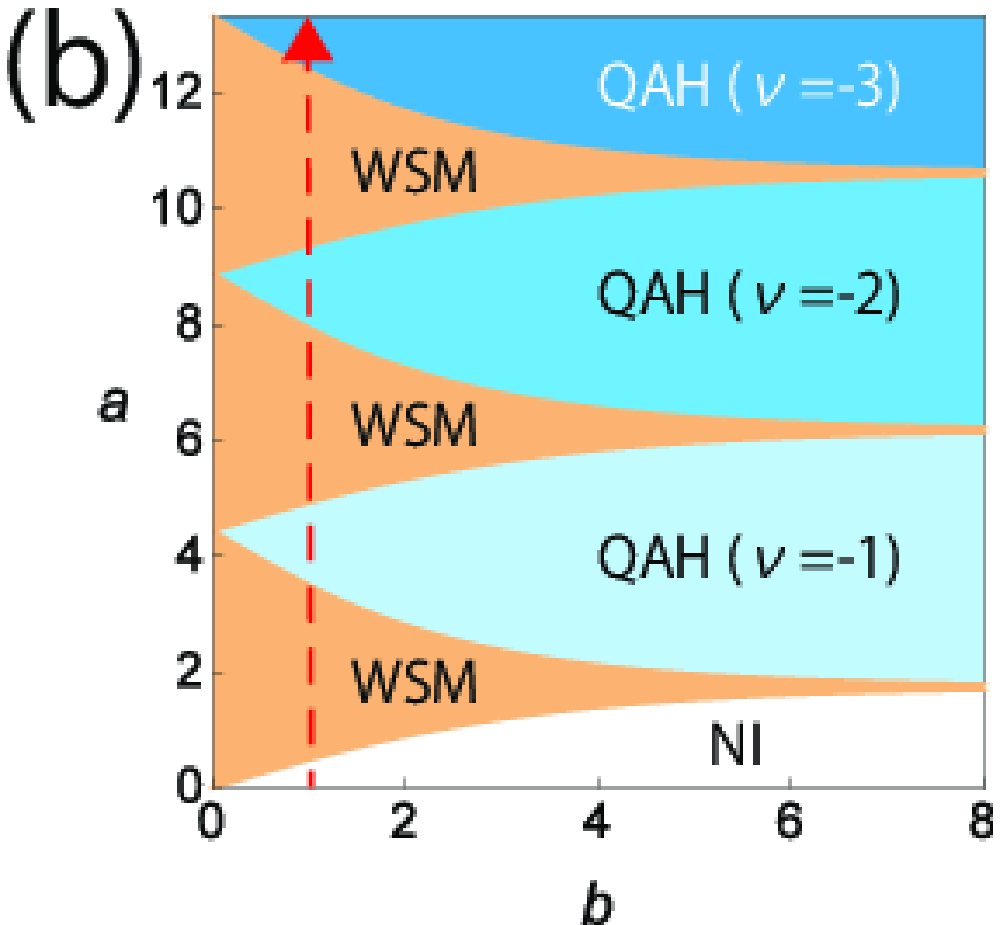}
\end{center}
\end{minipage}
\begin{minipage}[]{0.5\hsize}
\begin{center}
\includegraphics[width=4.0cm,height=4.0cm]{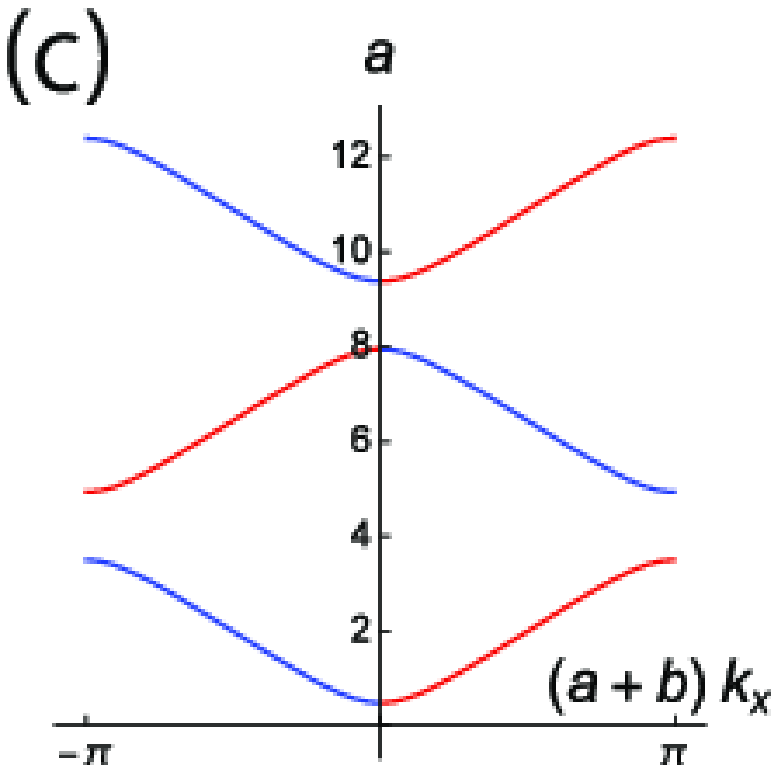}
\end{center}
\end{minipage}
\end{tabular}
\caption{\label{fig2} (Color online) (a) Schematic diagram of the multilayer of pattern B and the $x$ dependence of the parameter $m(x)$. (b) Phase diagram by changing the thicknesses $a$ and $b$ in the multilayer when $m_{+}=0.5$, $m_{-}=0.25$. The WSM phase is in the orange regions and the NI phase is in the white region. Each blue region is the QAH phase with different values of the Chern number $\nu$. (c) Positions of the Weyl nodes as the thickness $a$ is changed, with $b=1$ fixed and $m_{+}=0.5$, $m_{-}=0.25$. The Weyl nodes move along the $k_{x}$ axis. The red line represents the trajectory of a monopole and the blue line represents that of an anti-monopole. They correspond to the change of $a$ along the dashed line in (b).}
\end{figure}

In this section, we consider the WSM-NI multilayer of pattern B. The stacking direction is along the $x$ direction, which is parallel to the displacement vector ${\bm w}$ between the Weyl nodes. We set the parameter $m$ to be periodic as shown in Fig.~\ref{fig2} (a), and the Hamiltonian is given by
\begin{eqnarray}
&&H=\gamma\left[-\partial_{x}^2-m(x)\right]\sigma_{x}+v(k_{y}\sigma_{y}+k_{z}\sigma_{z}), \\
&&m\left(x\right)=\left\{ \begin{array}{ll}
m_{+}  & \left(0\leq x\leq a\right), \vspace{5pt}\\
-m_{-} & \left(-b\leq x\leq0\right),
\end{array}\right. \\
&&m\left(x+\left(a+b\right)\right)=m\left(x\right),
\label{eq101112}
\end{eqnarray}

\noindent where $m_{\pm}$ are positive constants.

We calculate the eigenstates and the energies in the similar way as in the Kr\"{o}nig-Penney model. The details of the calculation are given in Appendix \ref{subsec8}. In order to obtain the phase diagram as a function of $a$ and $b$, we examine whether the band gap closes. The band gap closes when
\begin{eqnarray}
&&\frac{m_{-}-m_{+}}{2\sqrt{m_{+}m_{-}}}\sin\sqrt{m_{+}}a\sinh\sqrt{m_{-}}b+\cos\sqrt{m_{+}}a\cosh\sqrt{m_{-}}b \nonumber\\
&&=\cos k_{x}\left(a+b\right),\hspace{3pt}k_{y}=k_{z}=0
\label{eq13}
\end{eqnarray}

\noindent is satisfied. Here, $k_{x}$ is the Bloch wave number along the $x$ axis. Solutions of Eq.~(\ref{eq13}) give positions of the Weyl nodes $\left(\pm k_{x},0,0\right)$. When Eq.~(\ref{eq13}) has (real) solutions of $k_{x}$, the system is in the WSM phase. This WSM phase is in the orange regions in Fig.~\ref{fig2} (b) for $m_{+}=0.5$ and $m_{-}=0.25$ as an example. 

In the other regions, the bulk is gapped. Because the time-reversal symmetry is broken, these regions with the bulk gap can be either the NI phase or the QAH phase. One can determine the phases in these regions in the following way. By continuity, the phases in the individual regions in the phase diagram can be easily determined by considering the limit $b\rightarrow\infty$ with fixed $a$, as shown in Appendix \ref{sec6}. In this limit, the system reduces to a thin slab of a WSM, and one can easily calculate its Chern number. We then obtain the phase diagram Fig.~\ref{fig2} (b). Here, the QAH phases are characterized by the Chern number $\nu$ within the constant $k_{z}$ plane. The values of the Chern number of two phases separated by the WSM phase are different by one.

We also investigate movement of the Weyl nodes in the wave-vector space when the thickness of the WSM layer is changed. Here, we fix the thickness of the NI layer as $b=1$, and we gradually increase the thickness of the WSM layer $a$. The positions of the Weyl nodes along the $k_{x}$ axis are shown in Fig.~\ref{fig2} (c) when $m_{+}=0.5$, $m_{-}=0.25$. First, when the multilayer enters the WSM phase as $a$ increases from zero, a pair of Weyl nodes is created at ${\bm k}={\bm 0}$. Then, the Weyl nodes continuously move along the $k_{x}$ axis as $a$ increases, and finally they are annihilated pairwise at the boundary of the Brillouin zone in the $x$ direction. Such pair creations and annihilations alternately occur as we increase $a$ further. Therefore, the phases with a bulk gap and the WSM phase alternately appear in the phase diagram of the multilayer. From the increment of the value of the Chern number by the increase of $a$, one can identify which trajectory is of a monopole or of an anti-monopole; the result is shown in Fig.~\ref{fig2} (c).

\section{\label{sec3} Multilayer from the lattice model}

In this section, we use the lattice model for a WSM proposed in Ref.~\onlinecite{0295-5075-97-6-67004} in order to study the WSM-NI multilayers. In Sec.~\ref{subsec4}, we introduce the lattice model for a WSM. By using this lattice model, we numerically calculate the band structure and phase diagrams for the multilayers with patterns A and B in Secs.~\ref{subsec5} and \ref{subsec6}, respectively.

\subsection{\label{subsec4} WSM from the lattice model}

\begin{figure}[]
\centering
\includegraphics[width=7.0cm,height=5.0cm]{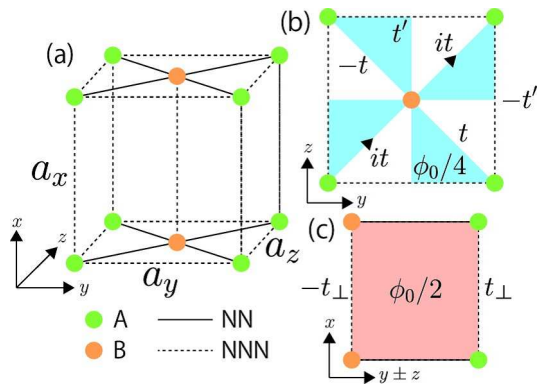}
\caption{\label{fig3} (Color online) (a) Three-dimensional (3D) unit cell of the lattice model. (b) Two-dimensional (2D) unit cell in the $y$-$z$ plane (c) Part of the unit cell along the $(110)$ and $(1\bar{1}0)$ planes, which contain the $A$ and $B$ sub-lattices. The solid and broken lines represent the nearest-neighbor hopping and the next-nearest-neighbor hopping, respectively. Colored regions in (b) and (c) are threaded by a magnetic flux: $\phi_{0}/4$ in the blue regions, and $\phi_{0}/2$ in the orange region, where $\phi_{0}=h/e$ is the magnetic flux quantum.}
\end{figure}

In this subsection, we review the lattice model proposed in Ref.~\onlinecite{0295-5075-97-6-67004}. It is a tight-binding model of spinless fermions on a lattice of stacked face-centered squares, shown in Fig.~\ref{fig3} (a). The unit cell of the lattice model includes sub-lattices $A$ and $B$. The nearest-neighbor hopping between the $A$ and $B$ sub-lattices has an amplitude $t$, and the next-nearest-neighbor hopping between the same sub-lattices has an amplitude $t^{\prime}$. Furthermore, the hopping amplitude along the stacking direction is denoted by $t_{\bot}$ between the same sub-lattices on adjacent layers. We also add an on-site energy for the $A$ and $B$ sub-lattices: $+\Delta$ for the $A$ sub-lattice and $-\Delta$ for the $B$ sub-lattice. The magnetic flux is added to the lattice with the gauge choice shown in Figs.~\ref{fig3} (b) and (c) with $\phi_{0}=h/e$, where $h$ is the Planck constant and $-e$ is the electron charge. The Hamiltonian of the lattice model is written as
\begin{eqnarray}
H&=&\left(2t\sin{\bm k}\cdot{\bm a}_{+}\right)\sigma_{x}+\left(2t\sin{\bm k}\cdot{\bm a}_{-}\right)\sigma_{y} \nonumber\\
&+&\left[\Delta-2t^{\prime}\left(\cos{\bm k}\cdot{\bm a}_{y}+\cos{\bm k}\cdot{\bm a}_{z}\right)+2t_{\bot}\cos{\bm k}\cdot{\bm a}_{x}\right]\sigma_{z}, \nonumber\\
\label{eq14}
\end{eqnarray}

\noindent where ${\bm a}_{x}$, ${\bm a}_{y}$ and ${\bm a}_{z}$ are primitive lattice vectors, and ${\bm a}_{\pm}=\left({\bm a}_{y}\pm{\bm a}_{z}\right)/2$. Here, we choose the coordinate axes as shown in Fig.~\ref{fig3} (a), which are different from Ref.~\onlinecite{0295-5075-97-6-67004}, in order to set the displacement vector ${\bm w}$ between the Weyl nodes to be along the $x$ axis in accordance with Fig.~\ref{fig0} (a). For simplicity, we set $\left|{\bm a}_{x}\right|=\left|{\bm a}_{y}\right|=\left|{\bm a}_{z}\right|=d$. The bulk energy is given by
\begin{eqnarray}
E&=&\pm2\left[t^2\left(\sin^2k_{+}d+\sin^2k_{-}d\right)\right. \nonumber\\
&+&\left.t_{\bot}^2\left(m_{1}-m_{2}\cos k_{+}d\cos k_{-}d+\cos k_{x}d\right)^2\right]^{1/2},
\label{eq15}
\end{eqnarray}

\noindent where $k_{\pm}=\left(k_{y}\pm k_{z}\right)/2$, $m_{1}=\Delta/t_{\bot}$, and $m_{2}=2t^{\prime}/t_{\bot}$.

The bulk gap closes when $\sin k_{+}d=\sin k_{-}d=0$ and $\cos k_{x}d=-\left(m_{1}-m_{2}\cos k_{+}d\cos k_{-}d\right)$. Within the region given by $m_{1}+m_{2}>1$, $|m_{1}-m_{2}|<1$, and $m_{1}$, $m_{2}>0$, the gap closes at $W_{\pm}=\left(\pm\frac{1}{d}\arccos\left(m_{2}-m_{1}\right),0,0\right)$ in the 3D wave-vector space. This region represents the WSM phase. On the other hand, when $m_{1}-m_{2}>1$ and $m_{1}$, $m_{2}>0$, the bulk is gapped, and the system is the NI phase.

In the following subsections \ref{subsec5} and \ref{subsec6}, we numerically calculate the bulk bands of the WSM-NI multilayers by use of the lattice model (\ref{eq14}) and compare the results with those in Secs.~\ref{subsec2} and \ref{subsec3}. In order to realize the WSM-NI multilayer using the lattice model (\ref{eq14}), the parameters $m_{1}$, $m_{2}$ are periodically modulated between those for a NI and those for a WSM. For simplicity, we fix the value of $m_{2}$ and change the value of $m_{1}$ to be $m_{\rm 1W}$ in the WSM layer and $m_{\rm 1N}$ in the NI layer, where the parameters $m_{\rm 1W}$ and $m_{\rm 1N}$ should meet the above conditions. Furthermore, we set $N_{a}$ and $N_{b}$ to be the numbers of the atomic layers within the WSM layer or the NI layer, respectively. Then, the thickness of the WSM layer and that of the NI layer are given by $a=N_{a}d$ and $b=N_{b}d$, respectively.

\subsection{\label{subsec5} Multilayer: Pattern A}

\begin{figure*}[]
\begin{tabular}{cc}
\begin{minipage}[]{0.35\hsize}
\begin{center}
\includegraphics[width=5.6cm,height=6.5cm]{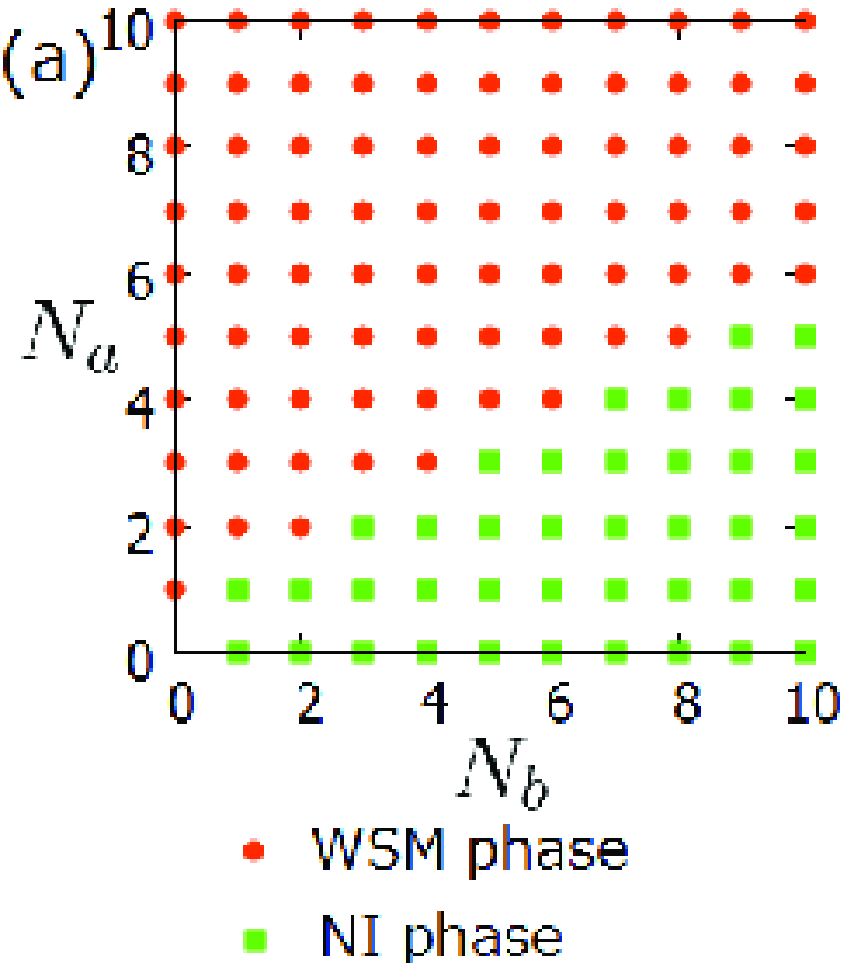}
\end{center}
\end{minipage}
\begin{minipage}[]{0.65\hsize}
\begin{center}
\includegraphics[width=10.6cm,height=6.5cm]{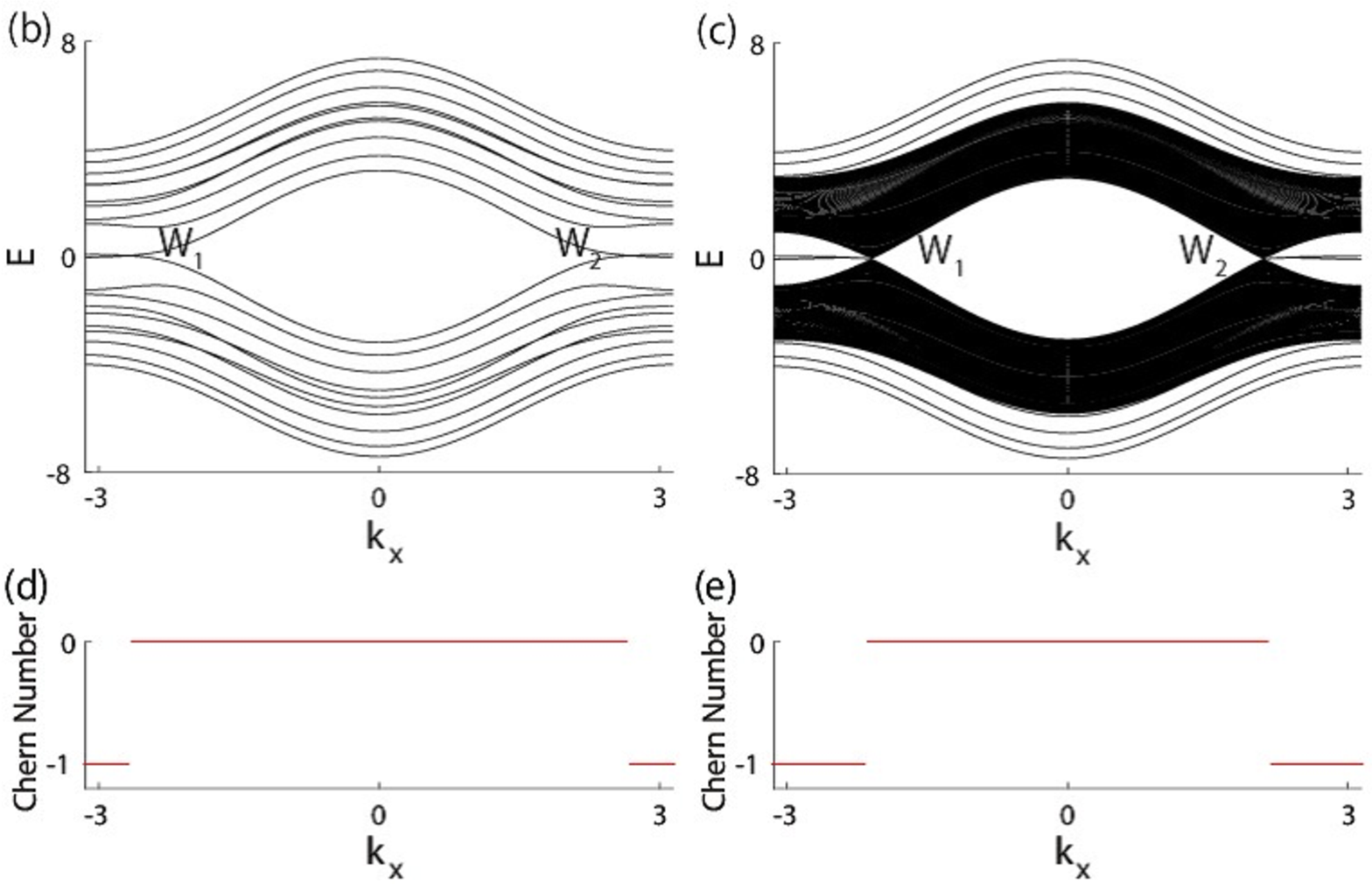}
\end{center}
\end{minipage}
\end{tabular}
\caption{\label{fig4} (Color online) (a) Phase diagram of the multilayer from the lattice model with the parameter values $m_{\rm 1W}=1.5$, $m_{\rm 1N}=2.5$, $m_{2}=1.0$. The orange circles represent the WSM phase and the green squares represent the NI phase. (b), (c) Bulk band structure and the Chern number of the multilayer with the parameter values $m_{\rm 1W}=1.5$, $m_{\rm 1N}=2.5$, $m_{2}=1.0$. The thicknesses of the layers are (b) $N_{a}=5$, $N_{b}=5$ and (c) $N_{a}=100$, $N_{b}=5$. The Weyl nodes move along the $k_{x}$ axis by changing the thickness of the WSM layer $a=N_{a}d$. (d), (e) Chern number of the multilayer with the parameter values $m_{\rm 1W}=1.5$, $m_{\rm 1N}=2.5$, $m_{2}=1.0$ and (d) $N_{a}=5$, $N_{b}=5$ and (e) $N_{a}=100$, $N_{b}=5$.}
\end{figure*}

In this subsection, we study the multilayer of the lattice model stacked along the $z$ direction, which is perpendicular to the direction of the displacement vector ${\bm w}$ between the two Weyl nodes. 

We numerically calculate the phase diagram of the multilayer from the lattice model with the parameter values $m_{\rm 1w}=1.5$, $m_{\rm 1N}=2.5$, $m_{2}=1.0$ as shown in Fig.~\ref{fig4} (a). In the phase diagram, the orange circles represent the WSM phase and the green squares represent the NI phase. This phase diagram qualitatively agrees with the result of Fig.~\ref{fig1} (a) from the effective model. Examples of the band structure are shown in Fig.~\ref{fig4} for $N_{a}=5$, $N_{b}=5$ in Fig.~\ref{fig4} (b) and for $N_{a}=100$, $N_{b}=5$ in Fig.~\ref{fig4} (c). In these cases, the Weyl nodes, ${\rm W}_{1}$ and ${\rm W}_{2}$, appear on the $k_{x}$ axis $(k_{y}=k_{z}=0)$. Comparing Figs.~\ref{fig4} (b) and \ref{fig4} (c), we find that the positions of the Weyl nodes move in the wave-vector space by changing the thickness of the WSM layer $a=N_{a}d$. It corresponds to the result in Eq.~(\ref{eq6}). Namely, as we decrease $N_{a}$ in Fig.~\ref{fig4} (a), the Weyl nodes gradually approach $k_{x}=\pm\pi$. Further decrease of $N_{a}$ causes a pair annihilation of Weyl nodes at $k_{x}=\pi$, driving the system into the NI phase. These Weyl nodes affect the Chern number on the $k_{y}k_{z}$ plane. The dependence of the Chern number of the wave number $k_{x}$ is shown in Figs.~\ref{fig4} (d) and (e) for $N_{a}=5$, $N_{b}=5$ and $N_{a}=5$, $N_{b}=100$, respectively. The Chern number on the plane $k_{x}={\rm const}.$ changes by $\pm1$ at the Weyl nodes. From these figures, we conclude that the Weyl node ${\rm W}_{1}$ has a monopole charge $+1$, and the Weyl node ${\rm W}_{2}$ has a monopole charge $-1$.

\begin{figure}[]
\begin{tabular}{cc}
\begin{minipage}[]{0.5\hsize}
\begin{center}
\includegraphics[width=5.8cm,height=3.5cm]{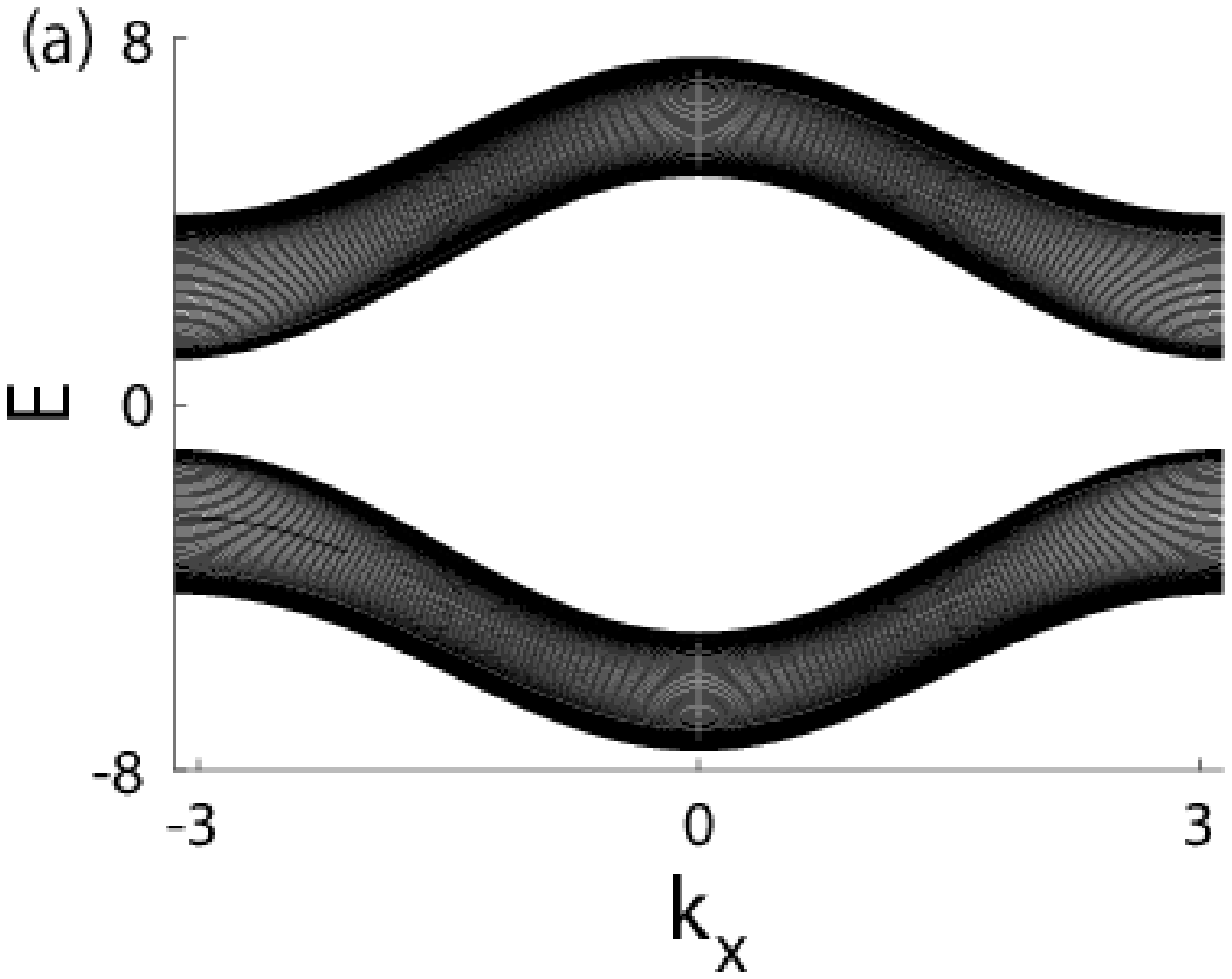}
\end{center}
\end{minipage}
\hfill
\begin{minipage}[]{0.5\hsize}
\begin{center}
\includegraphics[width=5.8cm,height=3.5cm]{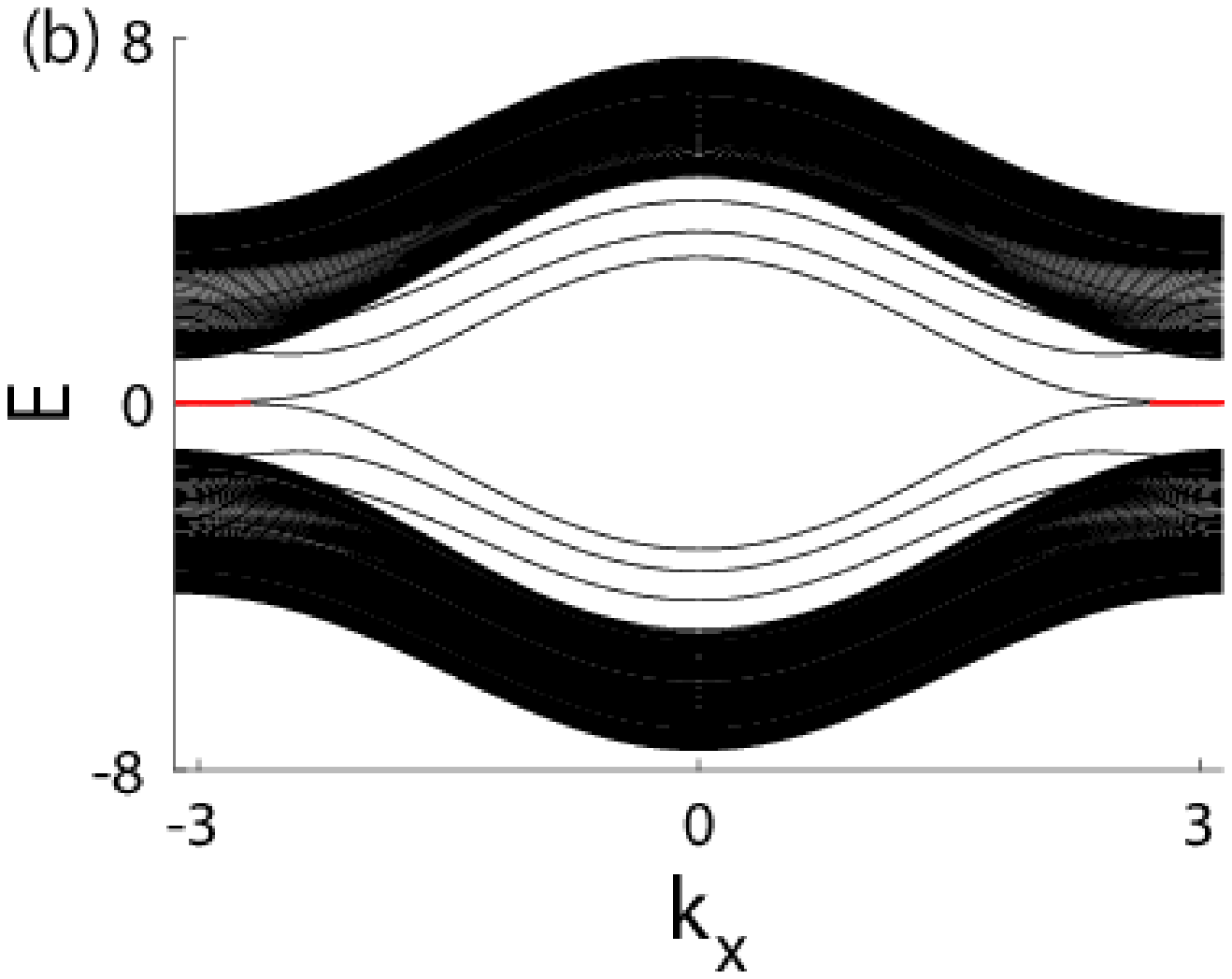}
\end{center}
\end{minipage}
\end{tabular}
\begin{center}
\includegraphics[width=7.3cm,height=4.5cm]{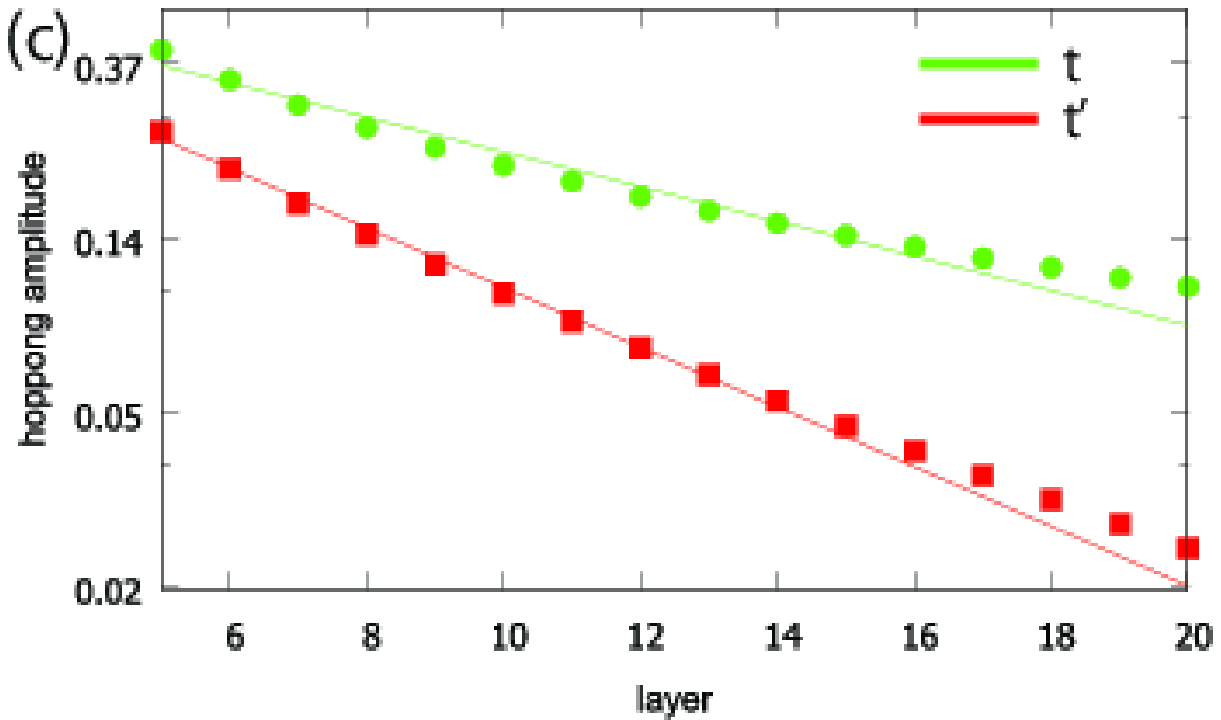}
\end{center}
\caption{\label{fig5} (a) Bulk band structure of the lattice model in the NI phase for the parameter values $N_{b}=100$, $m_{\rm 1N}=2.5$, $m_{2}=1.0$. (b) Band structure for the WSM-NI multilayer for the parameter values $N_{a}=5$, $N_{b}=100$, $m_{\rm 1W}=1.5$, $m_{\rm 1N}=2.5$, $m_{2}=1.0$. A band gap is almost zero as shown as red lines. (c) Hopping amplitudes across the layers as a function of the numbers of atomic layers, shown in a semi-logarithmic plot. The red squares represent the hopping amplitude $t^{\prime}$ through the NI layer with the parameter values $m_{\rm 1W}=1.5$, $m_{\rm 1N}=2.1$, $m_{2}=1.0$, $N_{a}=30$, $k_{y}=0$, $k_{x}=\pi$. The green circles represent the hopping amplitude $t$ through the WSM layer with the parameter values $m_{\rm 1W}=1.5$, $m_{\rm 1N}=2.5$, $m_{2}=1.0$, $N_{b}=30$, $k_{y}=0$, $k_{x}=2.1$. The red and the green lines show a fitting by exponential function.}
\end{figure}

In the above discussion, we found that the multilayer is in the WSM phase when the WSM layer is thick enough. On the other hand, when the NI layer is thick, it is in the NI phase. Nevertheless, for a sufficiently thick NI layer, the band structure, shown in Fig.~\ref{fig5} (b), looks quite different from that for the NI phase with uniform value of $m_{1}=m_{1N}$ as shown in Fig.~\ref{fig5} (a). Namely, it is similar to the band structure of the bulk of the NI [Fig.~\ref{fig5} (a)], but with additional bands around the Fermi energy. This behavior is caused by the Fermi arc which appears on the interfaces between the WSM layers and the NI layers. Namely, since the adjacent Fermi arcs are separated by the insulating layer, the states near the Fermi energy come from the Fermi arcs on the interfaces, with small hybridization. In Fig.~\ref{fig5} (b), there is a tiny gap throughout the whole Brillouin zone, and it represents a NI phase. In the following, we show this by a simple model.

Let $t$ and $t^{\prime}$ denote the hopping amplitudes between the adjacent Fermi arcs through the WSM layer and those through the NI layer, respectively. Furthermore, let $L\equiv a+b$ denote the period of the multilayer. An effective model describing hopping between the Fermi arcs is then written as
\begin{eqnarray}
H=\left( \begin{array}{cc}
0                             & t+t^{\prime}{\rm e}^{-ik_{z}L} \\
t+t^{\prime}{\rm e}^{ik_{z}L} & 0
\end{array}\right),
\label{eq16}
\end{eqnarray}

\noindent and its energy eigenvalues are given by 
\begin{equation}
E_{\pm}=\pm\sqrt{t^2+t^{\prime2}+2tt^{\prime}\cos k_{z}L}.
\label{eq17}
\end{equation}

\noindent The hopping amplitudes $t$, $t^{\prime}$ depend on $k_{x}$ and $k_{y}$. Physically, they depend on magnitudes of the band gaps of the two layers. For example, by increasing the thickness of the WSM layer or by increasing the band gap of the WSM layer, the hopping amplitude $t$ through this layer asymptotically becomes zero. The similar behavior is seen also for the NI layer. Namely, for the multilayer, $t^{\prime}\rightarrow0$ when we increase the thickness or the band gap of the NI layer. On the other hand, $t\rightarrow0$ when we increase thickness of the WSM layer at a wave number where the WSM layer has a gap. In order to confirm this scenario, we investigate magnitude of the hopping amplitudes $t$, $t^{\prime}$ as functions of the numbers of the layers in the multilayer using Eq.~(\ref{eq17}). For the hopping amplitude $t$ with the parameter values $m_{\rm 1W}=1.5$, $m_{\rm 1N}=2.1$, $m_{2}=1.0$, $N_{a}=30$, $k_{y}=0$, $k_{x}=\pi$, the result shown in Fig.~\ref{fig5} (c) fits well with exponentially decaying form $t\propto\exp\left(-n/\lambda\right)$ with $\lambda=5.83$. Similarly, the hopping amplitude $t^{\prime}$ with $m_{\rm 1W}=1.5$, $m_{\rm 1N}=2.5$, $m_{2}=1.0$, $N_{b}=30$, $k_{y}=0$, $k_{x}=2.1$ fits well with $t^{\prime}\propto\exp\left(-m/\lambda^{\prime}\right)$ with $\lambda^{\prime}=10.0$. Thus, this picture well explains the states close to $E=0$ in Fig.~\ref{fig5} (b). According to Eq.~(\ref{eq17}), the band gap closes when $t\left(k_{x},k_{y}\right)=\pm t^{\prime}\left(k_{x},k_{y}\right)$ within the WSM phase, and it is realized at some $k_{x}$ with $k_{y}=0$. Meanwhile, it is not satisfied in the NI phase and there appears a gap $2\left|t\pm t^{\prime}\right|$.

\subsection{\label{subsec6} Multilayer: Pattern B}

\begin{figure}[]
\begin{tabular}{cc}
\begin{minipage}[]{0.5\hsize}
\begin{center}
\includegraphics[width=4.2cm,height=6.0cm]{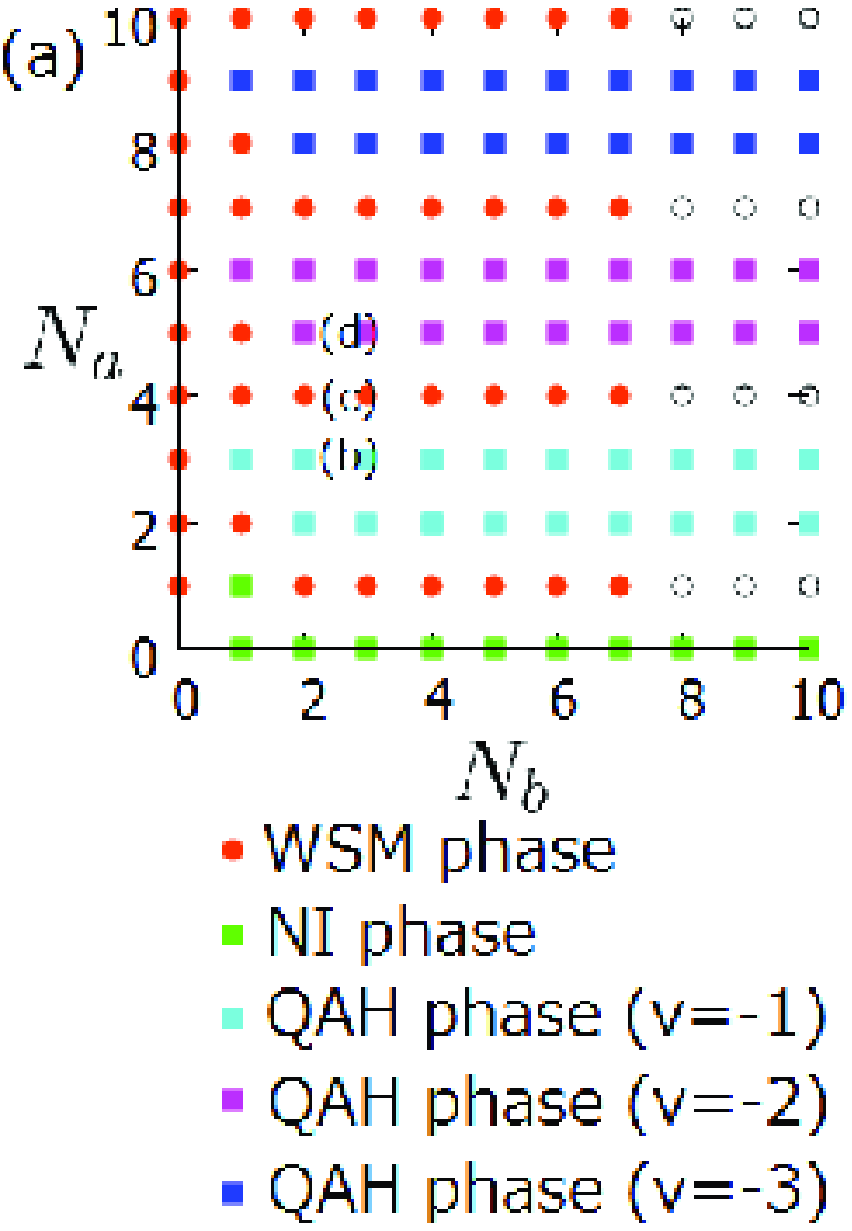}
\end{center}
\end{minipage}
\begin{minipage}[]{0.5\hsize}
\begin{center}
\includegraphics[width=4.5cm,height=6.0cm]{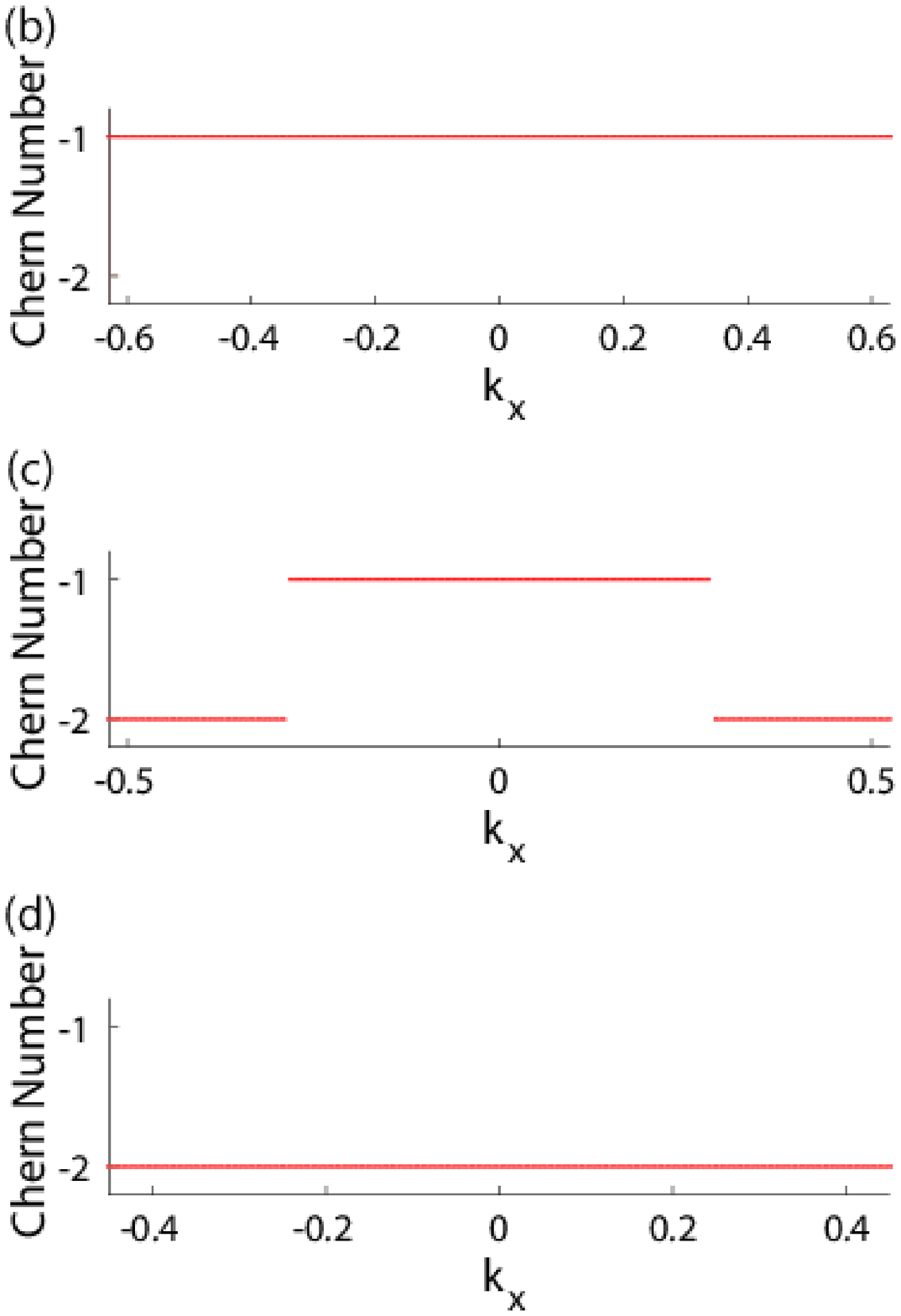}
\end{center}
\end{minipage}
\end{tabular}
\caption{\label{fig6} (Color online) (a) Phase diagram for the multilayer for the parameter values $m_{\rm 1W}=1.5$, $m_{\rm 1N}=2.1$, $m_{2}=1.0$. The circles represent the WSM phase, and the squares represent the phases with a bulk gap. The colors of the symbol refer as different phases. (b)-(d) Chern number as a function of $k_{x}$ with (b) $N_{a}=3$, $N_{b}=2$, (c) $N_{a}=4$, $N_{b}=2$, and (d) $N_{a}=5$, $N_{b}=2$. (b) Shows the QAH phase with the Chern number $-1$ for all $k_{x}$, (c) shows the WSM phase, and (d) shows the QAH phase with the Chern number $-2$ for all $k_{x}$.}
\end{figure}

We calculate the band structure for the multilayer of pattern B. We show our numerical results of the phase diagram and the Chern number with changing thicknesses of the WSM and the NI layers in Fig.~\ref{fig6}. In the phase diagram, the regions shown by the orange circles represent the WSM phase, and the regions shown by the squares represent phases with a bulk gap. Here, at the points represented by the white circles, we could not identify the phases numerically. The phases with a bulk gap (squares in Fig.~\ref{fig6}) are further classified as follows. Green represents the NI phase, and light blue, purple and blue represent the QAH phase with the Chern numbers $-1$, $-2$, and $-3$, respectively. The Chern number on the $k_{y}k_{z}$ plane as a function of $k_{x}$ for various values of the thickness of the WSM layer is shown in Fig.~\ref{fig6} for (b) $N_{a}=3$, $N_{b}=2$, (c) $N_{a}=4$, $N_{b}=2$, and (d) $N_{a}=5$, $N_{b}=2$. Figure.~\ref{fig6} (b) $\left(N_{a}=3,N_{b}=2\right)$ represents the QAH phase with $\nu=-1$ for any value of $k_{x}$. As the thickness of the WSM layer becomes larger as shown in Fig.~\ref{fig6} (c) $\left(N_{a}=4,N_{b}=2\right)$, the multilayer is in the WSM phase. Here, the Weyl nodes divide the $k_{x}$ axis into the regions of $\nu=-1$ and of $\nu=-2$. The phase transition between Figs.~\ref{fig6} (b) and (c) is attributed to a pair creation of Weyl nodes, and we interpret that it occurs at the boundary of the first Brillouin zone. As $N_{a}$ becomes larger, the Weyl nodes move further, until they annihilate pairwise. The multilayer is then in the QAH phase with $\nu=-2$ for all $k_{x}$ [Fig.~\ref{fig6} (d)]. Therefore, two bulk-insulating phases which sandwich the WSM phase have the Chern number different by $-1$. This conclusion agrees well with that from the effective model. Therefore, we have shown that the WSM-NI multilayer undergoes the phase transitions with various phases including the WSM phase and the QAH phase with various Chern numbers.

\section{\label{sec4} Summary and Discussion}

In this paper, we study behaviors of the WSM-NI multilayer using the effective model and the lattice model. In both models, the behaviors of the multilayer are different for different stacking directions. In the multilayer of pattern A, i.e., with the stacking direction perpendicular to the displacement vector ${\bm w}$ of the  Weyl nodes, the phase diagram is obtained as Fig.~\ref{fig1} (a), and the multilayer undergoes the phase transition from the NI phase to the WSM phase by increasing the thickness of the WSM layer. On the other hand, in pattern B, where the stacking direction is parallel to the displacement vector ${\bm w}$ of the Weyl nodes, the phase diagram is obtained as Fig.~\ref{fig2} (a). In the phase diagram, the WSM phase periodically appears as the thickness of the WSM layer varies, and two phases which sandwich the WSM phase are the QAH phases with different values of the Chern number. Agreement between the results from the effective model and those from the lattice model is surprisingly good.

At each interface in the multilayer with pattern A, Fermi-arc interface states exist; one may wonder whether they are safely incorporated in the calculation. In fact, in the detailed calculation in Appendix \ref{subsec7}, they are automatically included. When the wave vector $K$ of the wave function within the WSM layers [see Eq.~(\ref{eqa1})] is pure imaginary, the wave function is nothing but a Fermi-arc state, which follows from the properties of the effective model discussed in Ref.~\onlinecite{PhysRevB.89.235315}. The Fermi-arc interface states are included also in the calculation of the lattice model; indeed, the band structure in pattern A with relatively thick layers [e.g., Fig.\ref{fig5} (b)] can be understood from mixing of the Fermi arcs at the individual interfaces, with small hybridization between them. The exponential dependence of the hybridization as a function of the thickness [Fig.~\ref{fig5} (c)] confirms this scenario. Furthermore, we can apply the scenario of the mixing of the Fermi arcs not only in pattern A but also in pattern B. In pattern B, the Fermi arcs in the WSM layer are folded down into the first Brillouin zone due to the periodic stacking. It gives rise to multiple chiral surface states circulating along the $yz$ plane. As a result, the QAH phase appears on the phase diagrams [Figs.~\ref{fig2} (b) and \ref{fig6} (a)]. Therefore, appearance of the QAH phase can be regarded as a result of the hybridization between the Fermi arcs. Interestingly, the Chern number of the QAH phases can be large when we increase the thickness of the WSM layer $a$, and with this multilayer, one can design a QAH phase with large Chern number, which has been a long-standing issue in this field \cite{PhysRevLett.115.253901,PhysRevB.85.045445,PhysRevLett.111.136801,PhysRevLett.112.046801,PhysRevLett.113.113904,PhysRevA.91.043625}.

To the authors' knowledge, WSM-NI multilayers have not been studied previously. On the other hand, the WSM phase is found to appear in a TI-NI multilayer with magnetization \cite{PhysRevLett.107.127205} and in a stacked QAH layers system \cite{PhysRevB.94.235414}. In the phase diagrams of these two cases (TI-NI multilayer and stacked QAH layers system), the WSM phase appears between the NI and the QAH phases. By changing system parameters, one can go through NI-WSM-QAH phase transitions. Here, the WSM can be seen as an incomplete version of the QAH; namely, at some 2D slices in $\bm{k}$ space the Chern number is zero, while at others the Chern number is nontrivial. In this sense, our finding of an appearance of the QAH phase in the WSM-NI multilayer is nontrivial. As stated in the previous paragraph, the emergent QAH phase in pattern B is understood from the folding of the Brillouin zone by periodicity of the multilayer.

In this paper, we consider the WSM with broken time-reversal symmetry. There are a number of materials proposed as WSMs with broken time-reversal symmetry: pyrochlore iridates \cite{PhysRevB.83.205101,PhysRevB.84.075129,Y2Ir2O7}, CoTi$_2X$ ($X$=Si, Ge, Sn) \cite{Co2TiX}, ZrCo$_2$Sn \cite{magHeusler}, Hg$_{1-x-y}$Cd$_x$Mn$_y$Te in a magnetic field \cite{HgCdMnTe}, and Bi$_{0.97}$Sb$_{0.03}$ in a magnetic field \cite{Bi097Sb003}. The number of Weyl nodes, $N$, takes various even numbers in these WSMs. In WSMs with $N=2$, such as Hg$_{1-x-y}$Cd$_x$Mn$_y$Te under a magnetic field \cite{HgCdMnTe}, HgCr$_2$Se$_4$ \cite{PhysRevLett.107.186806,PhysRevLett.108.266802}, and Bi$_{0.97}$Sb$_{0.03}$ under a magnetic field \cite{Bi097Sb003}, there are only a pair of Weyl nodes around the Fermi energy. The present theory is directly applied, if the multilayer is either of the two patterns A and B; i.e., stacking vector $\bm{n}$, defined as a normal direction of individual layers, is either perpendicular or parallel to the displacement vector $\bm{w}$ between the two Weyl nodes. As we have discussed, in pattern A ($\bm{n}\perp\bm{w}$), the Weyl nodes approach each other by making a multilayer, and in pattern B ($\bm{n}\|\bm{w}$), the $\bm{k}$ space is folded into the Brillouin zone by the periodicity of the multilayer and becomes the WSM, the NI or the QAH phases depending on the thicknesses of the two layers.

In the WSM with $N=2$, one can think of general multilayers, where the two vectors $\bm{n}$ and $\bm{w}$ are neither parallel nor perpendicular. Such general cases can be understood by a combination of patterns A and B. In the direction perpendicular to the stacking vector $\bm{n}$, the Weyl nodes approach each other, while along the stacking vector $\bm{n}$, the $\bm{k}$ space is folded into the small Brillouin zone. Thus, the displacement vector between the Weyl nodes becomes gradually parallel as we increase $b/a$. As long as $\bm{n}$ is not parallel to the displacement vector between the Weyl nodes, the Weyl nodes have no chance of pair annihilation, and the system is in the WSM phase. Then, when $b/a$ exceeds a critical value, as is similar to pattern A, the displacement vector between the Weyl nodes becomes parallel to $\bm{n}$, and the multilayer may exhibit the QAH, the WSM, or the NI phases. The detailed behavior of the phases as a function of the thicknesses $a$ and $b$ depends on details of the system, and is left as a future work. 

In the other WSMs with broken time-reversal symmetry having $N>2$, the phase diagram for the multilayers can be discussed in a similar way, and the resulting phase diagram will become complicated. Nevertheless, based on the above analysis, one can say that possible phases are either a NI, a WSM, or a QAH phases, because the layered structure folds the Fermi arcs into the Brillouin zone of the multilayer. In general, a longer periodicity in the superlattice direction gives rise to larger Chern number, because the Fermi arcs are folded down many times. 

An extension to the WSM without the inversion symmetry is beyond the scope of this work, as the symmetry of the systems is very different. There are at least four Weyl nodes, and when the folding of the Brillouin zone gives rise to pair annihilations of all the Weyl nodes, it is expected to lead to either a NI or a weak TI phases. Nevertheless, the way how the Fermi arcs connect the four Weyl nodes depends on surface termination \cite{PhysRevB.89.235315}; therefore, a discussion from a viewpoint of Fermi arcs is not straightforward, and is left as a future work.

Thus far, we have studied multilayers with an infinite number of layers, and without any disorder. In reality, multilayers have a finite number of layers, and may have disorder. Strong disorder will eventually invalidate the physics discussed in this paper, because topological properties of a WSM phase rely on translational symmetry. Meanwhile, in a relatively clean system with small disorder, our results should remain valid. When the number of layers is sufficiently large, the multilayer behaves almost as a multilayer of infinite size, and the results discussed in this paper hold true. When the number of layers becomes smaller, the system is expected to exhibit a crossover between a 3D bulk multilayer to a 2D multilayer. This dimensional crossover gives a gap to the WSM phase, because the Weyl node is protected by topology in 3D wave-vector space, but not protected in 2D. On the other hand, the NI phase and the QAH phase are relatively robust. Therefore, in the phase diagram of the multilayer in pattern B, the WSM phase will acquire a gap, and, eventually, the phase diagram in the thinner multilayer would consist of only the NI phase and the QAH phase. How this dimensional crossover occurs is sensitive to the details of the multilayer, and is beyond the scope of this paper.

\begin{acknowledgments}
This work supported by Grant-in-Aid for Scientific Research (Grants No. 26287062 and No. 16K13834) by MEXT, Japan, by CREST, JST (Grant No. JPMJCR14F1), and by MEXT Elements Strategy Initiative to Form Core Research Center (TIES).
\end{acknowledgments}

\appendix

\section{\label{sec5} Calculations of eigenstates and energies for the WSM-NI multilayers}

\subsection{\label{subsec7} Multilayer: Pattern A}

Here, we explain the details of calculations of eigenstates and energy eigenvalues for the WSM-NI multilayer with Pattern A in Sec.~\ref{subsec2}. First, we write the wave function and the energy eigenvalue both in the WSM layer and in the NI layer, and connect the wave function at the boundaries. The wave function $\psi\left(z\right)$ in the WSM layer $\left(0<z<a\right)$ and in the NI layer $\left(-b<z<0\right)$ is given, respectively, by
\begin{eqnarray}
\psi\left(z\right)&=&A\left( \begin{array}{c}
E+vK                                 \vspace{3pt}\\
\gamma\left(k_{x}^2-m_{+}\right)+ivk_{y}         \\
\end{array}\right)\hspace{1.5pt}{\rm e}^{iKz} \nonumber\\
&+&B\left( \begin{array}{c}
E-vK                                 \vspace{3pt}\\
\gamma\left(k_{x}^2-m_{+}\right)+ivk_{y}         \\
\end{array}\right)\hspace{1.5pt}{\rm e}^{-iKz}\hspace{3pt}\left(0<z<a\right), \nonumber\\
\psi\left(z\right)&=&C\left( \begin{array}{c}
E-ivQ                                 \vspace{3pt}\\
\gamma\left(k_{x}^2+m_{-}\right)+ivk_{y}         \\
\end{array}\right)\hspace{1.5pt}{\rm e}^{Qz} \nonumber\\
&+&D\left( \begin{array}{c}
E+ivQ                                 \vspace{3pt}\\
\gamma\left(k_{x}^2+m_{-}\right)+ivk_{y}         \\
\end{array}\right)\hspace{1.5pt}{\rm e}^{-Qz}\hspace{3pt}\left(-b<z<0\right), \nonumber\\
\label{eqa1}
\end{eqnarray}

\noindent and the energy eigenvalue $E$ is written as
\begin{eqnarray}
E&=&\pm\sqrt{\gamma^2\left(k_{x}^2-m_{+}\right)^2+v^2\left(k_{y}^2+K^2\right)} \nonumber\\
&=&\pm\sqrt{\gamma^2\left(k_{x}^2+m_{-}\right)^2+v^2\left(k_{y}^2-Q^2\right)},
\label{eqa2}
\end{eqnarray}

\noindent where the $A$, $B$, $C$, $D$, $K$, and $Q$ are constants.

Hence, by continuity of the wave function and the Bloch condition, we obtain
\begin{eqnarray}
\left\{
\begin{array}{l}
\psi\left(+0\right)=\psi\left(-0\right), \vspace{1.5pt}\\
\psi\left(a-0\right)=\psi\left(-b+0\right)\hspace{1.5pt}{\rm e}^{ik_{z}\left(a+b\right)},
\end{array}\right.
\label{eqa3}
\end{eqnarray}

\noindent where $k_{z}$ is the Bloch wave number along the $z$ direction. By combining Eqs.~(\ref{eqa1}) - (\ref{eqa3}), we obtain the condition for the coefficients $A$, $B$, $C$, and $D$ to have non-trivial values:
\begin{eqnarray}
&&\frac{1}{KQ}\left[\frac{Q^2-K^2}{2}-\frac{\gamma^2\left(m_{+}+m_{-}\right)^2}{2v^2}\right]\sin Ka\sinh Qb \nonumber\\
&&+\cos Ka\cosh Qb=\cos k_{z}\left(a+b\right).
\label{eqa4}
\end{eqnarray}

\noindent Since $K$ and $Q$ are functions of the energy $E$ via Eq.~(\ref{eqa2}), Eq.~(\ref{eqa4}) determines the energy eigenvalues as functions of the Bloch wave vector ${\bm k}=\left(k_{x},k_{y},k_{z}\right)$.

Next, we derive the condition for closing of the band gap. We have set the Fermi energy to be $E_{F}=0$. Because the energy bands are symmetric with respect to $E=0$, the gap closing condition is obtained by setting $E=0$ in Eq.~(\ref{eqa4}). This yields Eq.~(\ref{eq6}) after a straightforward calculation.

\subsection{\label{subsec8} Multilayer: Pattern B}

In this subsection, we explain the details of calculations of eigenstates and energy eigenvalues for the WSM-NI multilayer with Pattern B in Sec.~\ref{subsec3}. First, we write the wave function and the energy eigenvalue in the WSM and the NI layers. In the WSM layer, the wave function is given by
\begin{eqnarray}
\psi\left(x\right)&=&\left( \begin{array}{c}
\gamma\left(-K_{1}^2+m_{+}\right)+ivk_{y} \vspace{3pt}\\
vk_{z}-E                               \\
\end{array}\right)\left(A_{1}\hspace{1.5pt}{\rm e}^{iK_{1}x}+B_{1}\hspace{1.5pt}{\rm e}^{-iK_{1}x}\right) \nonumber\\
&+&\left( \begin{array}{c}
\gamma\left(-K_{2}^2+m_{+}\right)+ivk_{y} \vspace{3pt}\\
vk_{z}-E                               \\
\end{array}\right)\left(C_{1}\hspace{1.5pt}{\rm e}^{iK_{2}x}+D_{1}\hspace{1.5pt}{\rm e}^{-iK_{2}x}\right) \nonumber\\
&&\left(0<x<a\right),
\label{eqa5}
\end{eqnarray}

\noindent and the energy eigenvalue is written as
\begin{eqnarray}
E&=&\pm\sqrt{\gamma^2\left(K_{1}^2-m_{+}\right)^2+v^2\left(k_{y}^2+k_{z}^2\right)} \nonumber\\
&=&\pm\sqrt{\gamma^2\left(K_{2}^2-m_{+}\right)^2+v^2\left(k_{y}^2+k_{z}^2\right)},
\label{eqa6}
\end{eqnarray}

\noindent where $A_{1}$, $B_{1}$, $C_{1}$, $D_{1}$, $K_{1}$, and $K_{2}$ are constants. The wave function and the energy eigenvalue in the NI layer $\left(-b<x<0\right)$ are given by replacing $m_{+}\rightarrow-m_{-}$, $K_{i}\rightarrow-iQ_{i}$ $\left(i=1,2\right)$, $A_{1}\rightarrow A_{2}$, $B_{1}\rightarrow B_{2}$, $C_{1}\rightarrow C_{2}$, and $D_{1}\rightarrow D_{2}$. Here, we select $K_{1}$ and $Q_{1}$ as
\begin{eqnarray}
\left\{
\begin{array}{l}
K_{1}^2=m_{+}+\frac{1}{\gamma}\sqrt{E^2-v^2\left(k_{y}^2+k_{z}^2\right)}, \vspace{1.5pt}\\
-Q_{1}^2=-m_{-}+\frac{1}{\gamma}\sqrt{E^2-v^2\left(k_{y}^2+k_{z}^2\right)}
\end{array}\right.
\label{eqa7}
\end{eqnarray}

\noindent and $K_{2}$ and $Q_{2}$ as
\begin{eqnarray}
\left\{
\begin{array}{l}
K_{2}^2=m_{+}-\frac{1}{\gamma}\sqrt{E^2-v^2\left(k_{y}^2+k_{z}^2\right)}, \vspace{1.5pt}\\
-Q_{2}^2=-m_{-}-\frac{1}{\gamma}\sqrt{E^2-v^2\left(k_{y}^2+k_{z}^2\right)}.
\end{array}\right.
\label{eqa8}
\end{eqnarray}

By continuity of the wave function and the Bloch condition we obtain
\begin{eqnarray}
\left\{
\begin{array}{l}
\psi\left(+0\right)=\psi\left(-0\right), \vspace{1.5pt}\\
\psi\left(a-0\right)=\psi\left(-b+0\right)\hspace{1.5pt}{\rm e}^{ik_{x}\left(a+b\right)}, \vspace{1.5pt}\\
\psi^{\prime}\left(+0\right)=\psi^{\prime}\left(-0\right), \vspace{1.5pt}\\
\psi^{\prime}\left(a-0\right)=\psi^{\prime}\left(-b+0\right)\hspace{1.5pt}{\rm e}^{ik_{x}\left(a+b\right)},
\end{array}\right.
\label{eqa9}
\end{eqnarray}

\noindent where $k_{x}$ is the Bloch wave number along the $x$ direction. Then, the condition for the coefficients $A_{i}$, $B_{i}$, $C_{i}$, and $D_{i}\hspace{5pt}(i=1,2)$ to have non-trivial values is
\begin{eqnarray}
&&\frac{Q_{i}^2-K_{i}^2}{2K_{i}Q_{i}}\sin K_{i}a\sinh Q_{i}b+\cos K_{i}a\cosh Q_{i}b \nonumber\\
&&=\cos k_{x}\left(a+b\right)\hspace{20pt}\left(i=1,2\right).
\label{eqa10}
\end{eqnarray}

\noindent Since $K_{i}$ and $Q_{i}$ $\left(i=1, 2\right)$ are functions of the energy $E$, Eq.~(\ref{eqa10}) determines the energy eigenvalues as functions of the Bloch wave vector ${\bm k}=\left(k_{x},k_{y},k_{z}\right)$.

We can derive the condition for closing of the band gap by setting $E=0$ in Eq.~(\ref{eqa10}), because the energy bands are symmetric with respect to $E=0$. This yields Eq.~(\ref{eq13}) after a straightforward calculation.

\section{\label{sec6} Calculation of the Chern number of a WSM slab from the effective model}

\begin{figure}[]
\centering
\includegraphics[width=5.0cm,height=4.0cm]{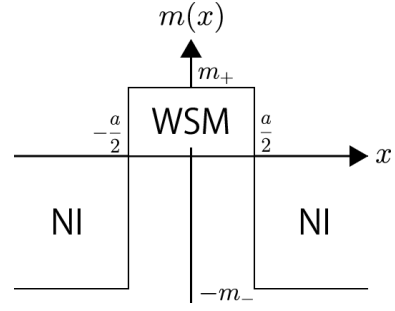}
\caption{\label{fig7} Value of $m\left(x\right)$ for the slab of the WSM with thickness $a$. It corresponds to the multilayer in the limit $b\rightarrow\infty$.}
\end{figure}

As seen from Fig.~\ref{fig2} (a), the phase diagram for pattern B in the region with a large value of $b$ consists mostly of bulk-insulating phases, separated from each other by the WSM phase, having a narrow region. These bulk-insulating phases might be the NI with zero Chern number or the QAH phase with nonzero Chern number. To calculate the Chern number for each phase, we note that the system in this region of the large value of $b$ is asymptotically a collection of WSM layers, separated far away from each other. Thus, the Chern number can be calculated by considering a single slab of a WSM with the thickness $a$, and this slab can be described within the effective model, when the parameter $m$ is as shown in Fig.~\ref{fig7}.

Let us calculate the Chern number for this WSM slab with the thickness $a$ in Fig.~\ref{fig7}. Here, the Chern number refers to that defined in the $k_{y}k_{z}$ plane. By the mirror symmetry with respect to the $yz$ plane, the mirror eigenvalues of the wave functions are either ${\cal M}=+1$ or ${\cal M}=-1$, yielding $\psi(-x)=\psi(x)$ and $\psi(-x)=-\psi(x)$, respectively. Namely, the wave functions are either symmetric or antisymmetric with respect to $x=0$.

Next, we write the wave functions in each region. The wave functions in the region $a/2<x$ can be written as
\begin{eqnarray}
\psi\left(x\right)&=&\left( \begin{array}{c}
\gamma\left(Q_{1}^2-m_{-}\right)+ivk_{y} \vspace{3pt}\\
vk_{z}-E                                             \\
\end{array}\right)\cdot A\hspace{1.5pt}{\rm e}^{-Q_{1}x} \nonumber\\
&+&\left( \begin{array}{c}
\gamma\left(Q_{2}^2-m_{-}\right)+ivk_{y} \vspace{3pt}\\
vk_{z}-E                                             \\
\end{array}\right)\cdot B\hspace{1.5pt}{\rm e}^{-Q_{2}x}\hspace{3pt}\left(\frac{a}{2}<x\right), \nonumber\\
\label{eqb1}
\end{eqnarray}

\noindent and the energy eigenvalues are given by
\begin{eqnarray}
E&=&\pm\sqrt{\gamma^2\left(-Q_{1}^2+m_{-}\right)^2+v^2\left(k_{y}^2+k_{z}^2\right)} \nonumber\\
&=&\pm\sqrt{\gamma^2\left(-Q_{2}^2+m_{-}\right)^2+v^2\left(k_{y}^2+k_{z}^2\right)}.
\label{eqb2}
\end{eqnarray}

\noindent Here, $A$, $B$, $Q_{1}$, and $Q_{2}$ are constants. Next, we write the wave function in the region $-a/2<x<a/2$. The wave functions with a mirror eigenvalue ${\cal M}=+1$ are given by
\begin{eqnarray}
\psi\left(x\right)&=&\left( \begin{array}{c}
\gamma\left(-K_{1}^2+m_{+}\right)+ivk_{y} \\
vk_{z}-E
\end{array}\right)\cdot C\cos K_{1}x \nonumber\\
&+&\left( \begin{array}{c}
\gamma\left(-K_{2}^2+m_{+}\right)+ivk_{y} \\
vk_{z}-E
\end{array}\right)\cdot D\cos K_{2}x \nonumber\\
&&\left(-\frac{a}{2}<x<\frac{a}{2}\right),
\label{eqb3}
\end{eqnarray}

\noindent and those with a mirror eigenvalue ${\cal M}=-1$ are given by
\begin{eqnarray}
\psi\left(x\right)&=&\left( \begin{array}{c}
\gamma\left(-K_{1}^2+m_{+}\right)+ivk_{y} \\
vk_{z}-E
\end{array}\right)\cdot C\sin K_{1}x \nonumber\\
&+&\left( \begin{array}{c}
\gamma\left(-K_{2}^2+m_{+}\right)+ivk_{y} \\
vk_{z}-E
\end{array}\right)\cdot D\sin K_{2}x \nonumber\\
&&\left(-\frac{a}{2}<x<\frac{a}{2}\right).
\label{eqb4}
\end{eqnarray}

\noindent Here, the energy eigenvalue is given by
\begin{eqnarray}
E&=&\pm\sqrt{\gamma^2\left(K_{1}^2-m_{+}\right)^2+v^2\left(k_{y}^2+k_{z}^2\right)} \nonumber\\
&=&\pm\sqrt{\gamma^2\left(K_{2}^2-m_{+}\right)^2+v^2\left(k_{y}^2+k_{z}^2\right)},
\label{eqb5}
\end{eqnarray}

\noindent and $C$, $D$, $K_{1}$, and $K_{2}$ are constants, and we select $K_{1}$, $Q_{1}$ as Eq.~(\ref{eqa7}) and $K_{2}$, $Q_{2}$ as Eq.~(\ref{eqa8}).

Now, by the continuity condition,
\begin{eqnarray}
\left\{
\begin{array}{l}
\psi\left(\frac{a}{2}+0\right)=\psi\left(\frac{a}{2}-0\right), \vspace{1.5pt}\\
\psi^{\prime}\left(\frac{a}{2}+0\right)=\psi^{\prime}\left(\frac{a}{2}-0\right), \vspace{1.5pt}\\
\end{array}\right.
\label{eqb6}
\end{eqnarray}

\noindent we derive a set of equations for the coefficients $A$, $B$, $C$, $D$, which eventually decouples to equations for $A$ and $C$, and those for $B$ and $D$. In either of these coupled equations, we can proceed in the similar way by changing the notation $K_{1}\rightarrow K$, $Q_{1}\rightarrow Q$ or $K_{2}\rightarrow K$, $Q_{2}\rightarrow Q$. Here, we adopt $B=D=0$ without losing generality The energy eigenvalues of the system are given by
\begin{equation}
E^2=\gamma^2M^2+v^2\left(k_{y}^2+k_{z}^2\right),
\label{eqb7}
\end{equation}

\noindent where $M$ is defined from $K$, $Q$ as
\begin{equation}
M\equiv K^2-m_{+}=-Q^2+m_{-}.
\label{eqb8}
\end{equation}

\noindent Finally, when the mirror eigenvalue is ${\cal M}=+1$, the condition for the existence of non-trivial solutions from Eq.~(\ref{eqb6}) is given by
\begin{equation}
\tan\frac{\sqrt{m_{+}+M}a}{2}=\sqrt{\frac{m_{-}-M}{m_{+}+M}}.
\label{eqb9}
\end{equation}

\noindent When a mirror eigenvalue is ${\cal M}=-1$, the condition is given by
\begin{equation}
\tan\left(\frac{\sqrt{m_{+}+M}a}{2}-\frac{\pi}{2}\right)=\sqrt{\frac{m_{-}-M}{m_{+}+M}}.
\label{eqb10}
\end{equation}

\noindent Thus, from Eqs.~(\ref{eqb9}) and (\ref{eqb10}), the thickness of the WSM layer $a$ determines the values of the parameter $M=M_{n}$ as
\begin{eqnarray}
&&\frac{\sqrt{m_{+}+M_{n}}a}{2}=\frac{n\pi}{2}+\arctan\sqrt{\frac{m_{-}-M_{n}}{m_{+}+M_{n}}}, \nonumber\\
&&\hspace{122pt}n=0, 1, 2, \cdots.
\label{eqb11}
\end{eqnarray}

\noindent For each value of $n$, the energy eigenvalues are given by Eq.~(\ref{eqb7}):
\begin{equation}
E^\pm_{n}\equiv\pm\sqrt{\gamma^2 M_{n}^2+v^2\left(k_{y}^2+k_{z}^2\right)}.
\label{eqb12}
\end{equation}

\begin{figure}[]
\centering
\includegraphics[width=4.4cm,height=5.8cm]{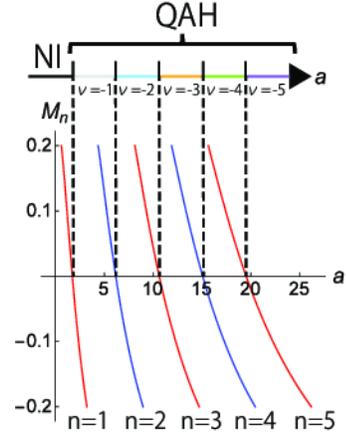}
\caption{\label{fig8} (Color online) Dependence of the parameter $M_{n}$ on the slab thickness $a$ determined by Eq.~(\ref{eqb9}) (red lines) and Eq.~(\ref{eqb10}) (blue lines) when $m_{+}=0.5$ and $m_{-}=-0.25$. The Chern number $\nu$ of the system changes by unity at the value of $a$ such that $M_{n}=0$.}
\end{figure}

\noindent Thus, $n$ and $\pm$ serve as a band index, and the corresponding eigenstates $\psi^{\pm}_{n}$ have a mirror eigenvalue ${\cal M}=\left(-1\right)^n$. In particular, from Eq.~(\ref{eqb12}), the system is gapless when $M_{n}=0$. We show the dependence of $M_{n}$ on the slab thickness $a$ in Fig.~\ref{fig8}. At the thickness $a$ such that $M_{n}=0$, the gap of the $n$th bands vanishes.

Next, we calculate the Chern number of the system from the wave functions calculated above. The Chern number on the $k_{y}k_{z}$ plane is defined as
\begin{eqnarray}
&&\nu=\sum_{n=0}^{\infty}\int_{\rm B.Z.}\frac{{\rm d} k_{y}\hspace{1.5pt}{\rm d} k_{z}}{2\pi}\hspace{1.5pt} \nonumber\\
&\times&i\left(\Braket{\frac{\partial\psi^{-}_{n{\bm k}}}{\partial k_{y}}|\frac{\partial\psi^{-}_{n{\bm k}}}{\partial k_{z}}}-\Braket{\frac{\partial\psi^{-}_{n{\bm k}}}{\partial k_{z}}|\frac{\partial\psi^{-}_{n{\bm k}}}{\partial k_{y}}}\right),
\label{eqb13}
\end{eqnarray}

\noindent where
\begin{eqnarray}
&&\Braket{\frac{\partial\psi^{-}_{n{\bm k}}}{\partial k_{y}}|\frac{\partial\psi^{-}_{n{\bm k}}}{\partial k_{z}}}-\Braket{\frac{\partial\psi^{-}_{n{\bm k}}}{\partial k_{z}}|\frac{\partial\psi^{-}_{n{\bm k}}}{\partial k_{y}}} \nonumber\\
&=&\int^{\infty}_{-\infty}{\rm d} x\hspace{1.5pt}\left(\frac{\partial\psi^{-\dag}_{n{\bm k}}(x)}{\partial k_{y}}\frac{\partial\psi^{-}_{n{\bm k}}(x)}{\partial k_{z}}-\frac{\partial\psi^{-\dag}_{n{\bm k}}(x)}{\partial k_{z}}\frac{\partial\psi^{-}_{n{\bm k}}(x)}{\partial k_{y}}\right), \nonumber\\
\label{eqb14}
\end{eqnarray}

\noindent and $\psi^{-}_{n{\bm k}}$ is the $n$th eigenstate at the Bloch wave vector ${\bm k}=\left(k_{y},k_{z}\right)$. The summation of the right-hand side of Eq.~(\ref{eqb13}) is taken over the states $\psi^{-}_{n{\bm k}}$ below the Fermi energy. Here, the integral is over the 2D Brillouin zone in the $k_{y}k_{z}$ plane. We note that the effective model describes the system only near ${\bm k}=0$, and we assume that the system itself has the periodicity of the Brillouin zone. As we increase the thickness $a$, the Chern number changes only when the band gap closes, i.e., $M_{n}=0$ for some $n$; this occurs at the intersections between the curves and the horizontal axis in Fig.~\ref{fig8}. Let $a=a_{\widetilde{n}}$ denote the thickness where $M_{\widetilde{n}}=0$ for some $\widetilde{n}$. The jump of the Chern number at $a=a_{\widetilde{n}}$ is calculated as
\begin{equation}
\nu\left(a_{\widetilde{n}}+0\right)-\nu\left(a_{\widetilde{n}}-0\right)=\left.\frac{1}{2}\Delta{\rm sgn}\left(M_{\widetilde{n}}\right)\right|_{a=a_{\widetilde{n}}}
\label{eqb15}
\end{equation}

\noindent where the right-hand side is $-1$ when $M_{\widetilde{n}}$ changes from positive to negative at $a=a_{\widetilde{n}}$, and is $+1$ when it changes from negative to positive as we increase $a_{\widetilde{n}}$. Therefore, from Fig.~\ref{fig8}, the Chern number changes by $-1$ as $a$ increases across $a_{\widetilde{n}}$. At $a=0$, the Chern number of the system is zero since the system is in the insulator phase. As $a$ gradually increases, the parameter $M_{n=0}$ becomes zero. The Chern number of the system then changes from zero to $-1$. As $a$ increases further, $M_{n=1}$ becomes zero and the Chern number of the system changes from $-1$ to $-2$. Thus, we have determined the Chern number for all the phases, and the insulator phase and the QAH phases appear as shown in Fig.~\ref{fig8}.

Let us go back to the multilayer. As the phase diagram of the slab (Fig.~\ref{fig8}) corresponds to the $b\rightarrow\infty$ limit of the multilayer, the resulting phase diagram of the multilayer is shown in Fig.~\ref{fig2} (a). As the thickness $a$ increases, the phases of the multilayer periodically change as NI $\rightarrow$ WSM $\rightarrow$ QAH $(\nu=-1)$ $\rightarrow$ WSM $\rightarrow$ QAH $(\nu=-2)$ $\rightarrow\cdots$.

\nocite{*}

\providecommand{\noopsort}[1]{}\providecommand{\singleletter}[1]{#1}%

\end{document}